\documentclass[11pt]{article}
\usepackage{amssymb,amsmath,amsfonts}
\usepackage{graphicx}
\usepackage{graphics}
\usepackage{eepic,epsfig}

\begin{document}

\title{Vacuum Currents for a Scalar Field in Models \\
with Compact Dimensions}
\author{Aram A. Saharian \\
%EndAName
\textit{Institute of Physics, Yerevan State University, }\\
\textit{1 Alex Manogian Street, 0025 Yerevan, Armenia }}
\maketitle

\begin{abstract}
This paper reviews the investigations on the vacuum expectation value of the
current density for a charged scalar field in spacetimes with toroidally
compactified spatial dimensions. As background geometries locally
Minkowskian (LM), locally de Sitter (LdS) and locally anti-de Sitter (LAdS)
spacetimes are considered. Along compact dimensions quasiperiodicity
conditions are imposed on the field operator and the presence of a constant
gauge field is assumed. The vacuum current has non-zero components only
along compact dimensions. Those components are periodic functions of the
magnetic flux enclosed by compact dimensions with the period equal to the
flux quantum. For LdS and LAdS geometries and for small values of the length
of a compact dimension, compared with the curvature radius, the leading term
in the expansion of the the vacuum current along that dimension coincides
with that for LM bulk. In this limit the dominant contribution to the mode
sum for the current density comes from the vacuum fluctuations with
wavelength smaller than the curvature radius and the influence of the
gravitational field is weak. The effects of the gravitational field are
essential for lengths of compact dimensions larger than the curvature
radius. In particular, instead of the exponential suppression of the current
density in LM bulk one can have power law decay in LdS and LAdS spacetimes.
\end{abstract}

\bigskip


\bigskip

\section{Introduction}

In a number of physical models the dynamics of the system is formulated in
background geometries having compact dimensions. The examples include
Kaluza-Klein type models with extra dimensions, string theories with
different types of compactifications of 6-dimensional internal subspace,
condensed matter systems like fullerenes, cylindrical nanotubes and toroidal
loops. The periodicity conditions imposed on dynamical variables along
compact dimensions are sources of a number of interesting effects, such as
the topological generation of mass, various mechanisms for symmetry breaking
and different kinds of instabilities (see, e.g., \cite{Ford79}-\cite{Quir03}%
). In quantum field theory the nontrivial spatial topology modifies the
spectrum of zero-point fluctuations of fields and as a consequence of that
the vacuum expectation values (VEVs) of physical observables are shifted by
an amount that depends on geometrical characteristics of the compact space.
This is the analog of the Casimir effect (for reviews see \cite{Most97}-\cite%
{Eliz94}), where the conditions imposed on constraining boundaries are
replaced by periodicity conditions, and it is known under the general name
of the topological Casimir effect. In Kaluza-Klein models this effect yields
an effective potential for the lengths of compact dimensions and can serve
as a compactification and stabilization mechanism for internal subspace
(Candelas-Weinberg mechanism \cite{Cand84}).

The most popular quantity in the investigations of the zero temperature
Casimir effect is the vacuum energy. Been an important global characteristic
of the vacuum state, it also determines the vacuum forces acting on
boundaries constraining the quantization volume. More detailed information
is contained in the local characteristics such as the vacuum expectation
value (VEV) of the energy-momentum tensor. The latter is a source of the
gravitational field in semiclassical Einstein equations and determines the
back-reaction of the quantum effects on the geometry of the spacetime. That
VEV of the energy-momentum tensor in the Casimir effect, in general, does
not obey the energy conditions in Hawking-Penrose singularity theorems and,
hence, is an interesting source of singularity free solutions for the
gravitational field. For charged fields, another important local
characteristic of the vacuum state is the expectation value of the current
density. It is a source of the electromagnetic field in Maxwell equations
and should be taken into account when considering the self-consistent
dynamics of the electromagnetic field.

Unlike to the energy-momentum tensor, in order to have nonzero vacuum
currents the parity symmetry of the model should be broken. That can be done
by introducing external fields or by imposing appropriate boundary or
periodicity conditions (in models with compact spatial dimensions). In
particular, the vacuum currents for charged scalar and fermionic fields in
locally Minkowski spacetime with toroidally compactified spatial dimensions
and in the presence of constant gauge fields have been investigated in \cite%
{Beze13,Bell10}. The parity symmetry in the corresponding models is broken
by an external gauge field or by quasiperiodic conditions along compact
dimensions with nontrivial phases. The results for a fermionic field in
(2+1)-dimensional spacetime have been applied to carbon nanotubes and
nanoloops. Those structures are obtained from planar graphene sheet by
appropriate identifications along a single or two spatial dimensions. In the
long wavelength approximation, the dynamics of electronic subsystem in
graphene is well described by an effective Dirac model with the Fermi
velocity of electrons appearing instead of the velocity of light (see, for
example, \cite{Gusy07,Cast09}). The corresponding quantum field theory lives
in spaces with topologies $R^{1}\times S^{1}$ and $S^{1}\times S^{1}$ for
nanotubes and nanoloops respectively. The appearance of the vacuum currents
along compact dimensions can be understood as a kind of the topological
Casimir effect. The combined influence of the boundary-induced and
topological Casimir effects on the vacuum currents have been studied in \cite%
{Bell15,Bell13} for scalar and fermionic fields on background of locally
flat spacetime with toral dimensions and in the presence of planar
boundaries.

The boundary conditions imposed on quantum fields serve as simplified models
for external fields and the Casimir effect can be considered as a vacuum
polarization sourced by those conditions. Another type of vacuum
polarization is induced by external gravitational fields. The combined
effects of those two sources on the VEV of the current density have been
investigated in \cite{Bell13dS} and \cite{Beze15,Bell17AdS} for locally de
Sitter (dS) and anti-de Sitter (AdS) spacetimes with a part of spatial
dimensions compactified to a torus. The high symmetry of these background
geometries allowed to find exact expressions for the current densities in
scalar and fermionic vacua. The effects of additional boundaries parallel to
the AdS boundary were studied in \cite{Bell15AdS}-\cite{Bell20AdS}. The
corresponding applications to higher dimensional braneworld models with
compact subspaces have been discussed.

The physical nature of the current densities considered in \cite%
{Beze13,Bell10} is similar to that for persistent currents in mesoscopic
rings \cite{Butt83}-\cite{Abra93}. Those currents are among the most
interesting manifestations of the Aharonov-Bohm effect and appear as a
result of phase coherence of the charge carriers, extended over the whole
mesoscopic ring. The persistent currents have been studied extensively in
the literature for electronic subsystems in different condensed matter
systems and for bosonic and fermionic atoms by making use of discrete or
continuum models (see, e.g., \cite{Rech07}-\cite{Chet22} and \cite{Levy90}-%
\cite{Pace22} for theoretical and experimental investigations, respectively,
and references therein). Their dependence on the geometry of ring is an
interesting direction of those investigations.

The present paper reviews the results of investigations of the vacuum
current densities for charged scalar fields in locally Minkowski, dS and AdS
spacetimes with toroidal subspaces. It is organized as follows. In Section %
\ref{sec:GenForm} we present the general formalism for the evaluation of the
current density for a charged scalar filed in models with compact
dimensions. The application of the formalism in the locally Minkowskian
background geometry is considered in Section \ref{sec:Mink}. The vacuum
currents for locally dS and AdS background spacetimes are studied in
Sections \ref{sec:dS} and \ref{sec:AdS}. The features of current densities
and comparative analysis for background geometries with zero, positive and
negative curvatures are discussed in Section \ref{sec:Feat}. The main
results are summarized in Section \ref{sec:Conc}. In appendix \ref{sec:App1}
we present the properties of the functions appearing in the expressions of
the current densities.

\section{General Formalism}

\label{sec:GenForm}

Consider a charged scalar field $\varphi (x)$ with the curvature coupling
parameter $\xi $ in background of $(D+1)$-dimensional spacetime described by
the line element $ds^{2}=g_{\mu \nu }(x)dx^{\mu }dx^{\nu }$. Here we use $%
x=(x^{0}=t,x^{1},\ldots ,x^{D})$ for the notation of spacetime points. In
the presence of a classical vector gauge field $A_{\mu }(x)$ the
corresponding action has the form%
\begin{equation}
S\left[ \varphi \right] =\frac{1}{2}\int d^{D+1}x\,\sqrt{|g|}\left[ g^{\mu
\nu }\left( D_{\mu }\varphi \right) \left( D_{\nu }\varphi \right) ^{\dagger
}+\left( \xi R+m^{2}\right) \varphi \varphi ^{\dagger }\right] .
\label{Action}
\end{equation}%
where $R$ is the Ricci scalar for the metric tensor $g_{\mu \nu }(x)$ and $%
D_{\mu }=\nabla _{\mu }+ieA_{\mu }$, with $\nabla _{\mu }$ being the related
covariant derivative operator. The most important special cases correspond
to the fields with minimal ($\xi =0$) and conformal ($\xi =\xi _{D}\equiv
(D-1)/(4D)$) couplings. The field equation obtained from (\ref{Action}) by
standard variational procedure reads
\begin{equation}
\left( g^{\mu \nu }D_{\mu }D_{\nu }+\xi R+m^{2}\right) \varphi (x)=0.
\label{Feq1}
\end{equation}%
We assume that a part of coordinates corresponding to the subspace $%
(x^{p+1},x^{p+2},\ldots ,x^{D})$ are compactified to circles of the lengths $%
(L_{p+1},L_{p+2},\ldots ,L_{D})$, respectively, and $0\leq x^{l}\leq L_{l}$,
$l=p+1,\ldots ,D$. In addition, the metric tensor is periodic along the
compact dimensions:
\begin{equation}
g_{\mu \nu }(t,x^{1},\ldots ,x^{p},\ldots ,x^{l}+L_{l},\ldots ,x^{D})=g_{\mu
\nu }(t,x^{1},\ldots ,x^{p},\ldots ,x^{l},\ldots ,x^{D}),  \label{metper}
\end{equation}%
where $p<l\leq D$. For the scalar and gauge fields less trivial
quasiperiodicity conditions are imposed
\begin{eqnarray}
\varphi (t,x^{1},\ldots ,x^{p},\ldots ,x^{l}+L_{l},\ldots ,x^{D})
&=&e^{i\alpha _{l}(x)}\varphi (t,x^{1},\ldots ,x^{p},\ldots ,x^{l},\ldots
,x^{D}),  \notag \\
A_{\mu }(t,x^{1},\ldots ,x^{p},\ldots ,x^{l}+L_{l},\ldots ,x^{D}) &=&A_{\mu
}(t,x^{1},\ldots ,x^{p},\ldots ,x^{l},\ldots ,x^{D})-(1/e)\nabla _{\mu
}\alpha _{l}(x),  \label{Qper}
\end{eqnarray}%
where the real functions $\alpha _{l}(x)$, $l=p+1,\ldots ,D$, are periodic
along the compact dimensions. With these conditions, the Lagrangian density
in (\ref{Action}) is periodic in the subspace $(x^{p+1},x^{p+2},\ldots
,x^{D})$. The gauge field is periodic up to a gauge transformation and this
type of quasiperiodic conditions are referred as C-periodic boundary
conditions (see, e.g., \cite{Hoof79,Gonz98,Anbe21}).

We are interested in the VEV of the current density
\begin{equation}
j_{\mu }=\frac{i}{2}e\left[ \{D_{\mu }\varphi ,\varphi ^{\dagger
}\}-\{D_{\mu }\varphi ,\varphi ^{\dagger }\}^{\dagger }\right] ,  \label{jmu}
\end{equation}%
where the figure brackets stand for the anticommutator. The VEVs of the
field bilinear combinations are expressed in terms of the Hadamard function $%
G(x,x^{\prime })$ defined as the VEV%
\begin{equation}
G(x,x^{\prime })=\left\langle \{\varphi (x),\varphi ^{\dagger }(x^{\prime
})\}\right\rangle ,  \label{GH}
\end{equation}%
where $\left\langle \cdots \right\rangle =\left\langle 0\right\vert \cdots
\left\vert 0\right\rangle $, with $\left\vert 0\right\rangle $ being the
vacuum state, stands for the VEV. In particular, for the VEV of the current
density we have%
\begin{equation}
\left\langle j_{\mu }(x)\right\rangle =\frac{ie}{2}\lim_{x^{\prime
}\rightarrow x}\left( D_{\mu }-g_{\mu }^{\quad \nu ^{\prime }}D_{\nu
^{\prime }}^{\ast }\right) G(x,x^{\prime }),  \label{jmuVev}
\end{equation}%
with $g_{\mu }^{\quad \nu ^{\prime }}$ being the bivector of parallel
displacement. We recall that the vector $\tilde{c}_{\mu }=g_{\mu }^{\quad
\nu ^{\prime }}c_{\nu ^{\prime }}$ is obtained by the parallel transport of $%
c_{\nu }$ from the point $x^{\prime }$ to the point $x$ along the geodesic
connecting those points.

The Hadamard function in (\ref{jmuVev}) can be evaluated in two different
ways: by solving the corresponding differential equation obtained from the
field equation or by using the complete set $\{\varphi _{\sigma
}^{(+)}(x),\varphi _{\sigma }^{(-)}(x)\}$ of the mode functions for the
field specified by the set of quantum numbers $\sigma $. We will follow the
second approach. For the field operator one has the expansion%
\begin{equation}
\varphi (x)=\sum_{\sigma }\left[ a_{\sigma }\varphi _{\sigma
}^{(+)}(x)+b_{\sigma }^{\dagger }\varphi _{\sigma }^{(-)}(x)\right] ,
\label{phiexp}
\end{equation}%
where $a_{\sigma }$ and $b_{\sigma }^{\dagger }$ are the annihilation and
creation operators. The symbol $\sum_{\sigma }$ stands for the summation
over the discrete quantum numbers in the set $\sigma $ and the integration
over the continuous ones. Substituting in the expression (\ref{GH}) for the
Hadamard function and by using the relations $a_{\sigma }\left\vert
0\right\rangle =b_{\sigma }\left\vert 0\right\rangle =0$, we get the mode sum%
\begin{equation}
G(x,x^{\prime })=\sum_{\sigma }\left[ \varphi _{\sigma }^{(+)}(x)\varphi
_{\sigma }^{(+)\ast }(x^{\prime })+\varphi _{\sigma }^{(-)}(x)\varphi
_{\sigma }^{(-)\ast }(x^{\prime })\right] .  \label{GHsum}
\end{equation}%
With this sum, the formal expression for the VEV of the current density
takes the form%
\begin{equation}
\left\langle j_{\mu }\right\rangle =-e\sum_{\sigma }\mathrm{Im}\left(
\varphi _{\sigma }^{(+)\ast }D_{\mu }\varphi _{\sigma }^{(+)}+\varphi
_{\sigma }^{(-)\ast }D_{\mu }\varphi _{\sigma }^{(-)}\right) .  \label{jms}
\end{equation}%
The expression on the right-hand side is divergent and a regularization with
the subsequent renormalization is required to obtain a finite result.

In the following sections we apply the presented general formalism for
special background geometries. Three cases will be considered: locally
Minkowski spacetime (LM), locally dS spacetime (LdS) and locally AdS
spacetime (LAdS). For all these geometries, the planar coordinates $\mathbf{x%
}_{q}=(x_{p+1},\ldots ,x_{D})$ will be used in the compact subspace. For the
classical gauge field we will assume a simple configuration $A_{\mu }=%
\mathrm{const}$ and in the periodicity conditions (\ref{Qper}) the phases $%
\alpha _{l}(x)=\alpha _{l}=\mathrm{const}$ will be taken. Special cases $%
\alpha _{l}=2\pi $ and $\alpha _{l}=\pi $ correspond to periodic and
antiperiodic fields and have been widely considered in the literature. The
constant gauge field is removed from the field equation for the scalar field
by the gauge transformation $\left( \varphi ,A_{\mu }\right) \rightarrow
\left( \varphi ^{\prime },A_{\mu }^{\prime }\right) $ given as
\begin{equation}
\varphi ^{\prime }(x)=\varphi (x)e^{ie\chi (x)},\;A_{\mu }^{\prime }=A_{\mu
}-\partial _{\mu }\chi (x),  \label{GT}
\end{equation}%
with the function $\chi (x)=A_{\mu }x^{\mu }$. For this function one gets $%
A_{\mu }^{\prime }=0$ and the field equation takes the form%
\begin{equation}
\left( g^{\mu \nu }\nabla _{\mu }\nabla _{\nu }+\xi R+m^{2}\right) \varphi
^{\prime }(x)=0.  \label{Feq1p}
\end{equation}%
The periodicity conditions for the new scalar field read%
\begin{equation}
\varphi ^{\prime }(t,x^{1},\ldots ,x^{p},\ldots ,x^{l}+L_{l},\ldots
,x^{D})=e^{i\tilde{\alpha}_{l}}\varphi ^{\prime }(t,x^{1},\ldots
,x^{p},\ldots ,x^{l},\ldots ,x^{D}),  \label{Qperp}
\end{equation}%
with new phases%
\begin{equation}
\tilde{\alpha}_{l}=\alpha _{l}+eA_{l}L_{l},  \label{alftilde}
\end{equation}%
for $l=p+1,\ldots ,D$. The physics is invariant under the gauge
transformation and we could expect that the components $A_{\mu }$, $\mu
=0,1,\ldots ,p$, will not appear in the expressions for physical quantities.
That is not the case for the components of the vector potential along the
compact dimensions. They will appear in the VEVs through the new phases (\ref%
{alftilde}). This is an Aharonov-Bohm type effect for a constant gauge field
in topologically nontrivial spaces. The further consideration will be
presented in terms of new fields $\left( \varphi ^{\prime },A_{\mu }^{\prime
}=0\right) $ omitting the primes.

\section{Locally Minkowski Spacetime with Toral Dimensions}

\label{sec:Mink}

We start the consideration with LM spacetime having the spatial topology $%
R^{p}\times T^{q}$, $q=D-p$, where the $q$-dimensional torus $T^{q}$
corresponds to the subspace with the compact coordinates $\mathbf{x}%
_{q}=(x^{p+1},x^{p+2},\ldots ,x^{D})$. For this geometry the metric in the
Cartesian coordinates is expressed as $g_{\mu \nu }=\eta _{\mu \nu }=\mathrm{%
diag}\left( 1,-1,\ldots ,-1\right) $. With this metric tensor one has $%
\nabla _{\mu }=\partial _{\mu }$. The topological Casimir effect in flat
spacetimes with toral dimensions has been widely considered in the
literature (see \cite{Most97,Eliz94}, \cite{Eliz89}-\cite{Hari22} and
references therein). As characteristics of the ground state for quantum
fields, the expectation values of the energy density and the stresses were
studied. The expectation value of the current density for scalar and
fermionic fields have been investigated in \cite%
{Beze13,Bell10,Bell15,Bell13,Bell14} at zero and finite temperatures. In
this section we review the results for the VEV\ of the current density for a
charged scalar field.

For the geometry under consideration the normalized scalar mode functions
are specified by the momentum $\mathbf{k}=\left( k_{1},k_{2},\ldots
,k_{D}\right) $ and are given by%
\begin{equation}
\varphi _{\mathbf{k}}^{(\pm )}(x)=(2^{p+1}\pi ^{p}V_{q}\omega _{\mathbf{k}%
})^{-1/2}e^{\mp i\omega _{\mathbf{k}}t+\mathbf{k}_{p}\cdot \mathbf{x}_{p}+%
\mathbf{k}_{q}\cdot \mathbf{x}_{q}},  \label{phik}
\end{equation}%
where $\mathbf{x}_{p}=(x^{1},\ldots ,x^{D})$ stands for the set of
coordinates in the uncompact subspace, $\omega _{\mathbf{k}}=\sqrt{\mathbf{k}%
^{2}+m^{2}}$ is the respective energy and $V_{q}=L_{p+1}....L_{D}$. For the
components of the momentum along uncompact dimensions, $\mathbf{k}%
_{p}=(k_{1},\ldots ,k_{p})$, we have $k_{l}\in (-\infty ,+\infty )$, $%
l=1,\ldots ,p$, whereas the eigenvalues along the compact dimensions, $%
\mathbf{k}_{q}=(k_{p+1},\ldots ,k_{D})$, are discretized by the periodicity
conditions (\ref{Qperp}):%
\begin{equation}
k_{l}=\frac{2\pi n_{l}+\tilde{\alpha}_{l}}{L_{l}},\quad n_{l}=0,\pm 1,\pm
2,\ldots .,  \label{kcomp}
\end{equation}%
with $l=p+1,...,D$. The integer part of the ratio $\tilde{\alpha}_{l}/(2\pi
) $ in the expressions of the VEVs is absorbed by shifting the integer
number $n_{l}$ in the corresponding summation and, hence, we can take $|%
\tilde{\alpha}_{l}|\leq \pi $ without loss of generality. Note that one has%
\begin{equation}
\varphi _{\mathbf{k}}^{(\pm )}(t,x^{1},\ldots ,x^{p},\ldots
,x^{l}-L_{l},\ldots ,x^{D})=e^{-i\tilde{\alpha}_{l}}\varphi _{\mathbf{k}%
}^{(\pm )}(t,x^{1},\ldots ,x^{p},\ldots ,x^{l},\ldots ,x^{D}),  \label{Qperm}
\end{equation}%
and hence for $\tilde{\alpha}_{l}\neq n\pi $, with $n$ being an integer, the
parity (P-) symmetry with respect to the reflection $x^{l}\rightarrow -x^{l}$
is broken by the corresponding quasiperiodicity condition. As it will be
seen below, the breaking of this inversion symmetry results in a non-zero
component of the current density along the $l$th compact dimension.

The Hadamard function is obtained from (\ref{GHsum}) with the modes (\ref%
{phik}). Because of the spacetime homogeneity the dependence on the
coordinates appears in the form of differences $\Delta t=t-t^{\prime }$, $%
\Delta \mathbf{x}_{p}\mathbf{=x}_{p}-\mathbf{x}_{p}^{\prime }$, and $\Delta
\mathbf{x}_{q}\mathbf{=x}_{q}-\mathbf{x}_{q}^{\prime }$. Inserting the modes
(\ref{phik}) in (\ref{GHsum}), one gets%
\begin{equation}
G(x,x^{\prime })=\frac{1}{V_{q}}\int \frac{d\mathbf{k}_{p}}{(2\pi )^{p}}%
\sum_{\mathbf{n}_{q}}e^{i\mathbf{k}_{p}\cdot \Delta \mathbf{x}_{p}+i\mathbf{k%
}_{q}\cdot \Delta \mathbf{x}_{q}}\frac{\cos (\omega _{\mathbf{k}}\Delta t)}{%
\omega _{\mathbf{k}}},  \label{GH2}
\end{equation}%
where $\mathbf{n}_{q}=(n_{p+1},\ldots ,n_{D})$. In the gauge under
consideration and for the geometry at hand the formula (\ref{jmuVev}) for
the current density takes the form
\begin{equation}
\left\langle j_{\mu }\right\rangle =\frac{ie}{2}\lim_{x^{\prime }\rightarrow
x}\left( \partial _{\mu }-\partial _{\mu ^{\prime }}\right) G(x,x^{\prime }).
\label{jmuM}
\end{equation}%
For $\mu =0,1,\ldots ,p$ the derivatives $\partial _{\mu }G(x,x^{\prime })$
and $\partial _{\mu ^{\prime }}G(x,x^{\prime })$ are odd functions of $%
\Delta x^{\mu }$ and the charge density and the components of the current
density along uncompact dimensions vanish, $\left\langle j_{\mu
}\right\rangle =0$ for $\mu =0,1,\ldots ,p$. Of course, we could expect that
from the problem symmetry under the reflections $x^{\mu }\rightarrow -x^{\mu
}$ along respective coordinates. In order to find the current density along
the $r$th compact dimension it is convenient to transform the corresponding
summation over $n_{r}$ in (\ref{GH2}). In order to do that we use a variant
of the Abel-Plana formula \cite{Bell09,Beze08}%
\begin{equation}
\frac{2\pi }{L_{r}}\sum_{n_{r}=-\infty }^{\infty }g(k_{r})f(\left\vert
k_{r}\right\vert )=\sum_{\lambda =\pm 1}\left[ \int_{0}^{\infty }dzg(\lambda
z)f(z)+i\int_{0}^{\infty }dz\,g(i\lambda z)\frac{f(iz)-f(-iz)}{%
e^{zL_{r}+i\lambda \tilde{\alpha}_{r}}-1}\right] ,  \label{APs}
\end{equation}%
with the functions $g(z)=e^{iz\Delta x^{r}}$ and $f(z)=\cos \left( \Delta t%
\sqrt{z^{2}+\omega _{p,q-1}^{2}}\right) /\sqrt{z^{2}+\omega _{p,q-1}^{2}}$,
where $\omega _{p,q-1}=\sqrt{\mathbf{k}_{p}^{2}+\mathbf{k}_{q-1}^{2}+m^{2}}$
and $\mathbf{k}_{q-1}=(k_{p+1},\ldots ,k_{r-1},k_{r+1},\ldots ,k_{D})$. By
making use of the expansion $1/(e^{y}-1)=\sum_{l=1}^{\infty }e^{-ly}$, the
integrals over $z$ and then over $\mathbf{k}_{p}$ are expressed in terms of
the Macdonald function $K_{\nu }(x)$. Introducing the notation $f_{\nu
}(x)=x^{-\nu }K_{\nu }(x)$, we get%
\begin{eqnarray}
G(x,x^{\prime }) &=&\frac{2L_{r}V_{q}^{-1}}{(2\pi )^{p/2+1}}%
\sum_{n_{r}=-\infty }^{+\infty }\sum_{\mathbf{n}_{q-1}}e^{in_{r}\tilde{\alpha%
}_{r}+i\mathbf{k}_{q-1}\cdot \Delta \mathbf{x}_{q-1}}\omega _{\mathbf{n}%
_{q-1}}^{p}  \notag \\
&&\times f_{\frac{p}{2}}\left( \omega _{\mathbf{n}_{q-1}}\sqrt{|\Delta
\mathbf{x}_{p}|^{2}+\left( \Delta x^{r}-n_{r}L_{r}\right) ^{2}-\left( \Delta
t\right) ^{2}}\right) ,  \label{GH3}
\end{eqnarray}%
where $\mathbf{n}_{q-1}=(n_{p+1},\ldots ,n_{r-1},n_{r+1},\ldots ,n_{D})$ and
$\omega _{\mathbf{n}_{q-1}}^{2}=\mathbf{k}_{q-1}^{2}+m^{2}$. The $n_{r}=0$
term here corresponds to the Hadamard function in the geometry with spatial
topology $R^{p+1}\times T^{q-1}$, where the $r$th dimension is
decompactified. The divergences in the coincidence limit $x^{\prime
}\rightarrow x$ are contained in that term only. The remaining part is
induced by the compactification of the coordinate $x^{r}$ and it is finite
in the coincidence limit. The latter property is related to the fact that
the compactification to a circle does not change the local geometry and,
hence, the structure of local divergences as well.

Plugging the Hadamard function (\ref{GH3}) in (\ref{jmuM}) and noting that
the term $n_{r}=0$ does not contribute to the component of the current
density along the $r$th compact dimension, for the corresponding
contravariant component one finds%
\begin{equation}
\left\langle j^{r}\right\rangle =\frac{2^{1-p/2}eL_{r}^{2}}{\pi ^{p/2+1}V_{q}%
}\sum_{n_{r}=1}^{\infty }n_{r}\sin (n_{r}\tilde{\alpha}_{r})\sum_{\mathbf{n}%
_{q-1}}\omega _{\mathbf{n}_{q-1}}^{p+2}f_{p/2+1}(n_{r}L_{r}\omega _{\mathbf{n%
}_{q-1}}).  \label{jr}
\end{equation}%
The specific features of the vacuum current will be discussed below in
Section \ref{sec:Feat}. As it has been already mentioned, for $\tilde{\alpha}%
_{r}=0,\pi $ the problem is symmetric with respect to the inversion $%
x^{r}\rightarrow -x^{r}$ and, as expected, the VEV $\left\langle
j^{r}\right\rangle $ vanishes. For those special values the contribution
from the right moving vacuum fluctuations with $k_{r}>0$ is cancelled by the
contribution coming from the left moving modes with $k_{r}<0$. For $\tilde{%
\alpha}_{r}=0$ there is also a zero mode with $k_{r}=0$ which does not
contribute to the current density along the $r$th dimension.

In the model with a single compact dimension $x^{D}$ one has $p=D-1$, $%
\omega _{\mathbf{n}_{q-1}}=m$ and the formula (\ref{jr}) is reduced to%
\begin{equation}
\left\langle j^{D}\right\rangle =\frac{4em^{D+1}L_{D}}{\left( 2\pi \right) ^{%
\frac{D+1}{2}}}\sum_{n=1}^{\infty }n\sin (n\tilde{\alpha}_{D})f_{\frac{D+1}{2%
}}(mnL_{D}).  \label{jrp1}
\end{equation}%
In particular, for a massless field we get%
\begin{equation}
\left\langle j^{D}\right\rangle =\frac{2e\Gamma \left( \frac{D+1}{2}\right)
}{\pi ^{\frac{D+1}{2}}L_{D}^{D}}\sum_{n=1}^{\infty }\frac{\sin (n\tilde{%
\alpha}_{D})}{n^{D}}.  \label{jrp1m0}
\end{equation}%
For odd values of $D$ the series is expressed in terms of the Bernoulli
polynomials $B_{n}(x)$ (see, e.g., \cite{Abra}) and we get%
\begin{equation}
\left\langle j^{D}\right\rangle =\frac{\left( -1\right) ^{\frac{D+1}{2}}\pi
^{\frac{D}{2}}e}{\Gamma \left( \frac{D}{2}+1\right) L_{D}^{D}}B_{D}\left(
\frac{\tilde{\alpha}_{D}}{2\pi }\right) ,  \label{jDp1odd}
\end{equation}%
for $0<\tilde{\alpha}_{D}<2\pi $. In particular, for $D=1$ and $D=3$ one
finds%
\begin{eqnarray}
\left\langle j^{D}\right\rangle &=&\frac{e}{L_{D}}\left( 1-\frac{\tilde{%
\alpha}_{D}}{\pi }\right) ,\;D=1,  \notag \\
\left\langle j^{D}\right\rangle &=&\frac{e\tilde{\alpha}_{D}}{6L_{D}^{3}}%
\left( 1-\frac{\tilde{\alpha}_{D}}{\pi }\right) \left( 2-\frac{\tilde{\alpha}%
_{D}}{\pi }\right) ,\;D=3.  \label{jD13}
\end{eqnarray}%
For $D\geq 2$ the current density is a continuous function of $\tilde{\alpha}%
_{D}$, whereas for $D=1$ the current density for a massless field is
discontinuous at $\tilde{\alpha}_{D}=2\pi n$ with integer $n$.

We could directly start from the mode sum formula (\ref{jms}) with $D_{\mu
}=\partial _{\mu }$. The substitution of the mode functions (\ref{phik})
leads to the expression%
\begin{equation}
\left\langle j^{r}\right\rangle =\frac{e}{V_{q}}\int \frac{d\mathbf{k}_{p}}{%
(2\pi )^{p}}\sum_{\mathbf{n}_{q}}\frac{k_{r}}{\omega _{\mathbf{k}}}=\frac{%
eL_{r}}{V_{q}}\frac{\partial }{\partial \tilde{\alpha}_{r}}\int \frac{d%
\mathbf{k}_{p}}{(2\pi )^{p}}\sum_{\mathbf{n}_{q}}\omega _{\mathbf{k}}.
\label{jr2}
\end{equation}%
Introducing the generalized zeta function $\zeta (s)$ in accordance with%
\begin{equation}
\zeta (s)=\frac{1}{V_{q}}\int \frac{d\mathbf{k}_{p}}{(2\pi )^{p}}\sum_{%
\mathbf{n}_{q}}\omega _{\mathbf{k}}^{-2s}=\frac{1}{V_{q}}\int \frac{d\mathbf{%
k}_{p}}{(2\pi )^{p}}\sum_{\mathbf{n}_{q}}\left( \mathbf{k}_{p}^{2}+\mathbf{k}%
_{q}^{2}+m^{2}\right) ^{-s},  \label{zetas}
\end{equation}%
the current density is written as
\begin{equation}
\left\langle j^{r}\right\rangle =eL_{r}\left. \frac{\partial }{\partial
\tilde{\alpha}_{r}}\zeta (s)\right\vert _{s=-1/2},  \label{jrzeta}
\end{equation}%
where $|_{s=-1/2}$ is understood in the sense of the analytical continuation
of the representation (\ref{zetas}) (for applications of the zeta function
technique in the investigations of the Casimir effect see, for example, \cite%
{Eliz94,Kirs02,Eliz12}). In order to realize the analytic continuation we
first integrate over the momentum in the uncompact subspace with the result%
\begin{equation}
\zeta (s)=\frac{\Gamma (s-p/2)}{2^{p}\pi ^{p/2}V_{q}\Gamma (s)}\sum_{\mathbf{%
n}_{q}}\left( \mathbf{k}_{q}^{2}+m^{2}\right) ^{p/2-s}.  \label{zetas2}
\end{equation}%
The application of the generalized Chowla-Selberg formula \cite{Eliz98} to
the multiple series in (\ref{zetas2}) gives%
\begin{equation}
\zeta (s)=\frac{m^{D-2s}}{\left( 4\pi \right) ^{\frac{D}{2}}\Gamma (s)}\left[
\Gamma \left( s-\frac{D}{2}\right) +2^{\frac{D}{2}+1-s}\sum_{\mathbf{n}%
_{q}}^{\prime }\cos \left( \mathbf{n}_{q}\cdot \mathbf{\tilde{\alpha}}%
_{q}\right) f_{\frac{D}{2}-s}(mg(\mathbf{L}_{q},\mathbf{n}_{q}))\right] ,
\label{zetas3}
\end{equation}%
where the prime on the summation sign means that the term with $\mathbf{n}%
_{q}=(0,0,\ldots ,0)$ should be excluded. Here we have introduced the $q$%
-component vector $\mathbf{\tilde{\alpha}}_{q}=(\tilde{\alpha}_{p+1},\tilde{%
\alpha}_{p+2},\ldots ,\tilde{\alpha}_{D})$ and the notation%
\begin{equation}
g(\mathbf{L}_{q},\mathbf{n}_{q})=\left(
\sum_{i=p+1}^{D}n_{i}^{2}L_{i}^{2}\right) ^{1/2}.  \label{gln}
\end{equation}%
The first term in the right-hand side of (\ref{zetas3}) corresponds to the
geometry without compact dimensions and it does not contribute to the
current density. The last term in (\ref{zetas3}) is finite at the physical
point and can be directly used in (\ref{jrzeta}) to obtain the following
expression for the current density
\begin{equation}
\left\langle j^{r}\right\rangle =\frac{2em^{D+1}L_{r}}{\left( 2\pi \right) ^{%
\frac{D+1}{2}}}\sum_{\mathbf{n}_{q}}^{\prime }n_{r}\sin \left( \mathbf{n}%
_{q}\cdot \mathbf{\tilde{\alpha}}_{q}\right) f_{\frac{D+1}{2}}(mg(\mathbf{L}%
_{q},\mathbf{n}_{q})).  \label{jr3}
\end{equation}%
In the special case of a single compact dimension $x^{D}$ this result
coincides with (\ref{jrp1}). Note that in the representation (\ref{jr3}) we
can make the replacement
\begin{equation}
n_{r}\sin \left( \mathbf{n}_{q}\cdot \mathbf{\tilde{\alpha}}_{q}\right)
\rightarrow n_{r}\sin (n_{r}\tilde{\alpha}_{r})\cos \left( \mathbf{n}%
_{q-1}\cdot \mathbf{\tilde{\alpha}}_{q-1}\right) .  \label{Replace}
\end{equation}%
The representation with this replacement is given in \cite{Beze13}. The
equivalence of two representations (\ref{jr}) and (\ref{jr3}) follows from
the relation%
\begin{equation}
\sum_{\mathbf{n}_{q-1}}\left( z^{2}+\mathbf{k}_{q-1}^{2}\right) ^{\frac{s+1}{%
2}}f_{\frac{s+1}{2}}(n_{r}L_{r}\sqrt{z^{2}+\mathbf{k}_{q-1}^{2}})=\frac{%
V_{q}z^{s+q}}{(2\pi )^{\frac{q-1}{2}}L_{r}}\sum_{\mathbf{n}_{q-1}}\cos (%
\mathbf{n}_{q-1}\cdot \boldsymbol{\tilde{\alpha}}_{q-1})f_{\frac{s+q}{2}}(zg(%
\mathbf{L}_{q},\mathbf{n}_{q})),  \label{RelSer}
\end{equation}%
with $s=p+1$ and $z=m$. This relation is proved in \cite{Bell09} by using
the Poisson's resummation formula. For a massless field, by using the
asymptotic for the modified Bessel function for small argument, one finds
\begin{equation}
\left\langle j^{r}\right\rangle =\frac{\Gamma \left( \frac{D+1}{2}\right) }{%
\pi ^{\frac{D+1}{2}}}eL_{r}\sum_{\mathbf{n}_{q}}^{\prime }\frac{n_{r}\sin
\left( \mathbf{n}_{q}\cdot \mathbf{\tilde{\alpha}}_{q}\right) }{g^{D+1}(%
\mathbf{L}_{q},\mathbf{n}_{q})}.  \label{jr3m0}
\end{equation}%
For a single compact dimension this formula coincides with (\ref{jrp1m0}).
Properties of the current density described by (\ref{jr}) and (\ref{jr3m0})
will be discussed in Section \ref{sec:Feat} below.

\section{Current Density in Locally dS Spacetime with Compact Dimensions}

\label{sec:dS}

In this section we consider $(D+1)$-dimensional locally dS spacetime with a
part of spatial dimensions compactified to $q$-dimensional torus in planar
(inflationary) coordinates. It is the solution of the Einstein field
equations with the positive cosmological constant $\Lambda $ as the only
source of the $(D+1)$-dimensional gravitation. The standard dS spacetime
(without compactification) is maximally symmetric and is one of the most
popular background geometries in gravity and in field theories. This has
several motivations. First of all, the high degree of symmetry allows to
have a relatively large number of exactly solvable problems. The
corresponding results shed light on the influence of the gravitational field
on various physical processes in more complicated curved backgrounds. In
accordance of the inflationary scenario, the dS spacetime approximates the
geometry of the early Universe and the investigation of the respective
effects is an important step to understand the dynamics of the Universe in
the postinflationary stage. In particular, the quantum fluctuations of
fields in the early dS phase of the expansion serve as seeds for large-scale
structure formation in the Universe. This is currently the most popular
mechanism for the formation of large-scale structures. Another motivation
for the importance of dS spacetime is conditioned by its role in the $%
\Lambda $CDM model for the cosmological expansion. In that model the
accelerated expansion of the Universe at recent epoch is sourced by a
positive cosmological constant and the dS spacetime appears as the future
attractor of the Universe expansion.

The explicit way to see the symmetries of the dS spacetime is its embedding
as a hyperboloid%
\begin{equation}
\left( Z^{0}\right) ^{2}-\left( Z^{1}\right) ^{2}-\cdots -\left(
Z^{D+1}\right) ^{2}=-a^{2},  \label{dShyp}
\end{equation}%
in $(D+2)$-dimensional Minkowski spacetime with the line element $ds_{D+2}^{%
\mathrm{(M)}2}=\left( dZ^{0}\right) ^{2}-\left( dZ^{1}\right) ^{2}-\cdots
-\left( dZ^{D+1}\right) ^{2}$. The parameter $a$ in (\ref{dShyp}) determines
the curvature radius of dS spacetime. Different coordinate systems have been
used to exclude an additional degree of freedom by using the relation (\ref%
{dShyp}). For the following discussion of the current density we will use
the planar coordinates $(\tau ,x^{1},x^{2},\ldots ,x^{D})$ which are
connected to the coordinates in the embedding spacetime by the relations
(see, for example, \cite{Grif09} in the case $D=3$)%
\begin{eqnarray}
Z^{i} &=&\frac{1}{2\tau }\left[ a^{2}+(-1)^{i}\left(
\sum_{l=1}^{D}(x^{l})^{2}-\tau ^{2}\right) \right] ,\;i=0,1,  \notag \\
Z^{l} &=&\frac{a}{\tau }x^{l-1},\;l=2,\ldots ,D+1.  \label{Zx}
\end{eqnarray}%
For the dS line element this gives $ds_{D+1}^{\mathrm{(dS)}2}=g_{\mu \nu
}(x)dx^{\mu }dx^{\nu }$, with the metric tensor%
\begin{equation}
g_{\mu \nu }(x)=\left( a/\tau \right) ^{2}\eta _{\mu \nu }.  \label{gdS}
\end{equation}%
In inflationary models a part of dS spacetime with the conformal time
coordinate in the range $-\infty <\tau <0$ is employed. For the
corresponding synchronous time coordinate $t$, $-\infty <t<+\infty $, one
has $t=-a\ln |\tau |/a$ and the line element is expressed as%
\begin{equation}
ds_{D+1}^{\mathrm{(dS)}2}=dt^{2}-\exp \left( \frac{2t}{\alpha }\right)
\sum_{i=1}^{D}(dx^{i})^{2}.  \label{dsdS}
\end{equation}%
The metric (\ref{gdS}) is the solution of Einstein's equations with positive
cosmological constant $\Lambda =\frac{D(D-1)}{2a^{2}}$ as the only source of
the gravitational field.

The hyperbolic embedding (\ref{dShyp}) in the locally Minkowski spacetime
with the line element $ds_{D+2}^{\mathrm{(M)}2}$ works equally well for LdS
spacetime with a $q$-dimensional toroidal subspace covered by the
coordinates $(x^{p+1},x^{p+2},\ldots ,x^{D})$. The coordinate $x^{l}$, $%
l=p+1,\ldots ,D$, varies in the range $0\leq x^{l}\leq L_{l}$. In this case,
the subspace with the coordinates $(Z^{p+2},\ldots ,Z^{D+2})$ is
compactified to a torus. The length of the compact dimension $Z^{l+1}$, $%
l=p+1,\ldots ,D$, is given by $L_{\mathrm{(p)}l}=aL_{l}/\eta =e^{t/a}L_{l}$,
with $\eta =|\tau |$. Note that $L_{\mathrm{(p)}l}$ is the proper length of
the compact dimension $x^{l}$ in the LdS spacetime.

For LdS spacetime with the metric tensor (\ref{gdS}) and compact subspace $%
\mathbf{x}_{q}=(x^{p+1},x^{p+2},\ldots ,x^{D})$, the scalar mode functions
can be presented in the form $\varphi _{\sigma }^{(+)}(x)=\eta ^{D/2}g_{\nu
}(k\eta )e^{i\mathbf{k}_{p}\cdot \mathbf{x}_{p}\mathbf{+ik}_{q}\cdot \mathbf{%
x}_{q}}$, where $g_{\nu }(y)$ is a cylinder function of the order%
\begin{equation}
\nu =\left[ \frac{D^{2}}{4}-D(D+1)\xi -a^{2}m^{2}\right] ^{1/2}.  \label{nju}
\end{equation}%
The eigenvalues of the momentum components along compact dimensions are
given by (\ref{kcomp}). Different choices of the function $g_{\nu }(y)$
correspond to different vacuum states for a scalar field in dS spacetime.
Here we will investigate the current density in the Bunch-Davies vacuum
state \cite{Bunc78}.

\subsection{Hadamard Function}

For the Bunch-Davies vacuum state the normalized scalar mode functions are
specified by $\sigma =\mathbf{k}=(\mathbf{k}_{p},\mathbf{k}_{q})$ and are
expressed as%
\begin{equation}
\left\{
\begin{array}{l}
\varphi _{\mathbf{k}}^{(+)}(x) \\
\varphi _{\mathbf{k}}^{(-)}(x)%
\end{array}%
\right\} =\left( \frac{a^{1-D}e^{i(\nu -\nu ^{\ast })\pi /2}}{2^{p+2}\pi
^{p-1}V_{q}}\right) ^{\frac{1}{2}}e^{i\mathbf{k}_{p}\cdot \mathbf{x}_{p}%
\mathbf{+ik}_{q}\cdot \mathbf{x}_{q}}\eta ^{D/2}\left\{
\begin{array}{l}
H_{\nu }^{(1)}(k\eta ) \\
H_{\nu ^{\ast }}^{(2)}(k\eta )%
\end{array}%
\right\} ,  \label{phidS}
\end{equation}%
where $H_{\nu }^{(1,2)}(y)$ are the Hankel functions and the star stands for
the complex conjugate. Note that, depending on the curvature coupling
parameter and on the mass of the field quanta, the order of the Hankel
functions can be either nonnegative real or purely imaginary. With the modes
(\ref{phidS}), the mode sum for the Hadamard function reads%
\begin{equation}
G(x,x^{\prime })=\frac{\left( \eta \eta ^{\prime }\right) ^{\frac{D}{2}}e^{i%
\frac{\pi }{2}(\nu -\nu ^{\ast })}}{2^{p+2}\pi ^{p-1}V_{q}a^{D-1}}\int d%
\mathbf{k}_{p}\,\sum_{\mathbf{n}_{q}}e^{i\mathbf{k}_{p}\cdot \Delta \mathbf{x%
}_{p}+i\mathbf{k}_{q}\cdot \Delta \mathbf{x}_{q}}[H_{\nu }^{(1)}(k\eta
)H_{\nu ^{\ast }}^{(2)}(k\eta ^{\prime })+H_{\nu }^{(1)}(k\eta ^{\prime
})H_{\nu ^{\ast }}^{(2)}(k\eta )].  \label{GHdS}
\end{equation}%
Applying to the sum over $n_{r}$ the summation formula (\ref{APs}) and
assuming that ${\mathrm{Re}\,\nu <1}$, one finds the representation%
\begin{eqnarray}
G(x,x^{\prime }) &=&\frac{4\left( \eta \eta ^{\prime }\right) ^{\frac{D}{2}%
}L_{r}}{\left( 2\pi \right) ^{\frac{p+3}{2}}V_{q}a^{D-1}}\int_{0}^{\infty
}d\lambda \,\lambda \left[ I_{-\nu }(\eta \lambda )K_{\nu }(\eta ^{\prime
}\lambda )+K_{\nu }(\eta \lambda )I_{\nu }(\eta ^{\prime }\lambda )\right]
\sum_{\mathbf{n}_{q}}e^{in_{r}\alpha _{r}}  \notag \\
&&\times e^{i\mathbf{k}_{q-1}\cdot \Delta \mathbf{x}_{q-1}}\left( \lambda
^{2}+\mathbf{k}_{q-1}^{2}\right) ^{\frac{p-1}{2}}f_{\frac{p-1}{2}}(\sqrt{%
\lambda ^{2}+\mathbf{k}_{q-1}^{2}}\sqrt{|\Delta \mathbf{x}_{p}|^{2}+\left(
\Delta x^{r}-n_{r}L_{r}\right) ^{2}}),  \label{GHdS2}
\end{eqnarray}%
where $I_{\nu }(y)$ and $K_{\nu }(y)$ are the modified Bessel functions.
Similar to the case of the Minkowski bulk, the contribution coming from the
term $n_{r}=0$ gives the Hadamard function for the geometry where the $r$th
coordinate is decompactified (spatial topology $R^{p+1}\times T^{q-1}$). The
effects of the compactification of that coordinate are included in the part
with $n_{r}\neq 0$.

For a conformally coupled massless field we have $\nu =1/2$ and
\begin{equation}
I_{-\nu }(\eta \lambda )K_{\nu }(\eta ^{\prime }\lambda )+K_{\nu }(\eta
\lambda )I_{\nu }(\eta ^{\prime }\lambda )=\frac{\cosh \left( \lambda \Delta
\eta \right) }{\lambda \sqrt{\eta \eta ^{\prime }}}.  \label{IKcc}
\end{equation}%
With this function, the integral over $z$ in (\ref{GHdS2}) is evaluated by
using the formula from \cite{PrudV2} and we get%
\begin{equation}
G(x,x^{\prime })=\left( \frac{\eta \eta ^{\prime }}{a^{2}}\right) ^{\frac{D-1%
}{2}}G_{\mathrm{M}}(x,x^{\prime }),  \label{GHdScc}
\end{equation}%
where the Minkowskian Hadamard function $G_{\mathrm{M}}(x,x^{\prime })$ is
given by (\ref{GH3}) with $m=0$. For a conformally coupled massless field
this is the standard relation between two conformally related geometries.

\subsection{Vacuum Current}

For the components of the current density along uncompact dimensions $x^{\mu
}$, $\mu =1,2,\ldots ,p$, one has $\partial _{\mu }G(x,x^{\prime })\propto
g_{\mu \alpha }\Delta x^{\alpha }$ and the corresponding expectation values
vanish. By using the properties of the modified Bessel functions we can see
that%
\begin{equation}
\lim_{x^{\prime }\rightarrow x}(\partial _{0}-\partial _{0}^{\prime
})\left\{ \left( \eta \eta ^{\prime }\right) ^{\frac{D}{2}}\left[ I_{-\nu
}(\eta \lambda )K_{\nu }(\eta ^{\prime }\lambda )+K_{\nu }(\eta \lambda
)I_{\nu }(\eta ^{\prime }\lambda )\right] \right\} =0,  \label{der0}
\end{equation}%
and, hence, the charge density vanishes as well. In order to find the
component of the current density along the $r$th compact dimension we use
the representation (\ref{GHdS2}) for the Hadamard function in combination
with (\ref{jmuVev}) where $D_{\mu }=\partial _{\mu }$. In (\ref{GHdS2}), the
derivative of the term $n_{r}=0$ with respect to $x^{r}$ is an odd function
of $\Delta x^{r}$ and vanishes in the coincidence limit. As it has been
mentioned before, that term corresponds to the Hadamard function in the
geometry with uncompactified $x^{r}$ and the corresponding current density
vanishes by the symmetry. The part in (\ref{GHdS2}) induced by the
compactification of the direction $x^{r}$ (the terms with $n_{r}\neq 0$) is
finite in the coincidence limit and that limit can be directly taken in the
expression for the VEV. This gives the following expression for the
contravariant component \cite{Bell13dS}:
\begin{eqnarray}
\left\langle j^{r}\right\rangle &=&\frac{8ea(\eta /a)^{D+2}}{\left( 2\pi
\right) ^{\frac{p+3}{2}}V_{q}}L_{r}^{2}\int_{0}^{\infty }d\lambda \,\lambda %
\left[ I_{-\nu }(\eta \lambda )+I_{\nu }(\eta \lambda )\right] K_{\nu }(\eta
\lambda )  \notag \\
&&\times \sum_{n_{r}=1}^{\infty }n_{r}\sin (n_{r}\tilde{\alpha}_{r})\sum_{%
\mathbf{n}_{q-1}}\left( \lambda ^{2}+\mathbf{k}_{q-1}^{2}\right) ^{\frac{p+1%
}{2}}f_{\frac{p+1}{2}}(n_{r}L_{r}\sqrt{\lambda ^{2}+\mathbf{k}_{q-1}^{2}}).
\label{jdS}
\end{eqnarray}%
An alternative representation for the current density is obtained by
applying the formula (\ref{RelSer}) with $s=p$. This gives%
\begin{eqnarray}
\left\langle j^{r}\right\rangle &=&\frac{4eL_{r}}{\left( 2\pi \right) ^{%
\frac{D}{2}+1}a^{D+1}}\int_{0}^{\infty }du\,u^{D+1}\left[ I_{-\nu
}(u)+I_{\nu }(u)\right] K_{\nu }(u)  \notag \\
&&\times \sum_{\mathbf{n}_{q}}n_{r}\sin (\mathbf{n}_{q}\cdot \boldsymbol{%
\tilde{\alpha}}_{q})f_{\frac{D}{2}}(ug(\mathbf{L}_{q},\mathbf{n}_{q})/\eta ).
\label{jdS2}
\end{eqnarray}%
The integral in the right-hand side is evaluated by using the formula%
\begin{equation}
\int_{0}^{\infty }dz\,z^{\frac{D}{2}+1}\left[ I_{\nu }(z)+I_{-\nu }(z)\right]
K_{\nu }(z)K_{\frac{D}{2}}(bz)=\frac{\sqrt{2\pi }}{4}b^{\frac{D}{2}}p_{\nu -%
\frac{1}{2}}^{-\frac{D+1}{2}}\left( \frac{b^{2}}{2}-1\right) ,  \label{Intf2}
\end{equation}%
where we use the notation%
\begin{equation}
p_{\alpha }^{-\mu }(u)=\Gamma (\mu -\alpha )\Gamma \left( \mu +\alpha
+1\right) \frac{P_{\alpha }^{-\mu }(u)}{\left( u^{2}-1\right) ^{\frac{\mu }{2%
}}},  \label{pmu}
\end{equation}%
with $P_{\nu }^{\mu }(u)$ being the associated Legendre function of the
first kind. The expression of the function $p_{\alpha }^{-\mu }(u)$ in terms
of the hypergeometric function is given in Appendix \ref{sec:App1}. The
result (\ref{Intf2}) is obtained from the integral involving the product $%
I_{\nu }(z)K_{\nu }(z)K_{\frac{D}{2}}(bz)$ and given in \cite{PrudV2}. That
integral is expressed in terms of the sum of two hypergeometric functions.
The contribution of the second function is canceled in evaluating the
integral (\ref{Intf2}). Then, we express the hypergeometric function in
terms of the associated Legendre functions. By taking into account (\ref%
{Intf2}) in (\ref{jdS2}) the current density is expressed as
\begin{equation}
\left\langle j^{r}\right\rangle =\frac{eL_{r}}{\left( 2\pi \right) ^{\frac{%
D+1}{2}}a^{D+1}}\sum_{\mathbf{n}_{q}}n_{r}\sin (\mathbf{n}_{q}\cdot
\boldsymbol{\tilde{\alpha}}_{q})p_{\nu -\frac{1}{2}}^{-\frac{D+1}{2}}\left(
\frac{g^{2}(\mathbf{L}_{q},\mathbf{n}_{q})}{2\eta ^{2}}-1\right) .
\label{jdS3}
\end{equation}%
In both the formulas (\ref{jdS2}) and (\ref{jdS3}) we can make the
replacement (\ref{Replace}). Note that one has the property $p_{\nu -\frac{1%
}{2}}^{-\frac{D+1}{2}}\left( u\right) =p_{-\nu -\frac{1}{2}}^{-\frac{D+1}{2}%
}\left( u\right) $ and the expression on the right-hand side is real for
both real and purely imaginary values of $\nu $.

\section{AdS Spacetime with Compact Dimensions}

\label{sec:AdS}

Now we turn to the LAdS spacetime with a part of spatial dimensions
compactified to a torus. It obeys the $(D+1)$-dimensional Einstein equations
with the negative cosmological constant $\Lambda $. The usual AdS spacetime
is maximally symmetric and appears as the ground state in supergravity and
in string theories. That was the reason for the early interest in the AdS
physics. The interest to those investigations was further increased related
to two exciting developments in modern theoretical physics. The first one,
dubbed as AdS/CFT correspondence (see, e.g., \cite{Ahar00}-\cite{Nats15}),
establishes duality between the supergravity and string theory on the AdS
bulk and conformal field theory (CFT) on its boundary. This duality is a
unique way to investigate strong coupling effect in one theory by mapping it
on the dual theory. A number of examples can be found in the literature,
including those with applications in condensed matter physics. The second
development, with the AdS spacetime as a background geometry, corresponds to
braneworld models of the Randall-Sundrum type \cite{Maar10} with large extra
dimensions. In the corresponding setup the standard model fields are
localized in a 4-dimensional hypersurface (brane) on background of higher
dimensional AdS spacetime. The braneworld models provide a geometrical
solution to the hierarchy problem between the electroweak and Planck energy
scales and naturally arise in the string/M theory context. They present a
novel setting in considerations of various phenomenological and cosmological
issues, in particular, the generation of cosmological constant localized on
the brane.

By analogy of the dS bulk, it is convenient to visualize the AdS spacetime
as a hyperboloid%
\begin{equation}
\left( Z^{0}\right) ^{2}-\sum_{i=1}^{D}\left( Z^{i}\right) ^{2}+\left(
Z^{D+1}\right) ^{2}=-a^{2},  \label{hypAdS}
\end{equation}%
where the line element of the embedding $(D+2)$-dimensional flat spacetime
is given by $ds_{D+2}^{2}=\left( dZ^{0}\right) ^{2}-\sum_{i=1}^{D}\left(
dZ^{i}\right) ^{2}+\left( dZ^{D+1}\right) ^{2}$. The Poincar\'{e}
coordinates $(t,x^{1}=z,x^{2},\ldots ,x^{D})$ are introduced by the relations%
\begin{eqnarray}
Z^{i} &=&\frac{1}{2z}\left[ a^{2}+(-1)^{i}\left(
\sum_{l=1}^{D}(x^{l})^{2}-t^{2}\right) \right] ,\;i=0,1,  \notag \\
Z^{l} &=&\frac{a}{z}x^{l},\;Z^{D+1}=\frac{a}{z}t,\;l=2,\ldots ,D.
\label{ZPuan}
\end{eqnarray}%
In those coordinates the metric tensor of the AdS spacetime is expressed as
\begin{equation}
g_{\mu \nu }=\frac{a^{2}}{z^{2}}\eta _{\mu \nu },  \label{gAdS}
\end{equation}%
with $0\leq z<\infty $. The hypersurfaces $z=0$ and $z=\infty $ present the
AdS boundary and the horizon, respectively. The proper distance along the
direction $x^{1}$ is measured by the coordinate $y=a\ln (z/a)$, $-\infty
<y<+\infty $, in terms of which the line element is written as%
\begin{equation}
ds^{2}=\exp \left( -\frac{2y}{a}\right) \left[ dt^{2}-%
\sum_{l=2}^{D}(dx^{l})^{2}\right] -dy^{2}.  \label{dsAdS}
\end{equation}%
Here we consider the LAdS geometry with the coordinates $(x^{p+1},x^{p+2},%
\ldots ,x^{D})$ compactified to a torus as described in section \ref%
{sec:GenForm}. It can be embedded as a hyperboloid (\ref{hypAdS}) in the
spacetime with coordinates $(Z^{0},Z^{1},\ldots ,Z^{D+1})$, where the
coordinate $Z^{l}$, $l=p+1,\ldots ,D$, is compactified to a circle with the
length $L_{\mathrm{(p)}l}=aL_{l}/z=e^{-y/a}L_{l}$. The latter is the proper
length for the compact dimension in LAdS. It is exponentially small near the
horizon.

The scalar mode functions in the coordinates corresponding to the metric
tensor (\ref{gAdS}) and obeying the periodicity conditions (\ref{Qperp}) are
written in the form $\varphi _{\sigma }^{(\pm )}(x)=e^{\mp i\omega t+i%
\mathbf{k}_{p-1}\mathbf{x}_{p-1}+\mathbf{k}_{q}\mathbf{x}_{q}}f(z)$, with $%
\mathbf{x}_{p-1}=(x^{2},\ldots ,x^{p})$, $\mathbf{k}_{p-1}=(k_{2},\ldots
,k_{p})$. The equation for the function $f(z)$ is obtained from the field
equation. The corresponding solution is presented as $z^{D/2}\left[
c_{1}J_{\nu _{+}}(\lambda z)+c_{2}Y_{\nu _{+}}(\lambda z)\right] $, where $%
J_{\nu }(\lambda z)$ and $Y_{\nu }(\lambda z)$ are the Bessel and Neumann
functions, $\lambda ^{2}=\omega ^{2}-\mathbf{k}_{p-1}^{2}-\mathbf{k}_{q}^{2}$%
, and%
\begin{equation}
\nu _{+}=\left[ \frac{D^{2}}{4}-D(D+1)\xi +a^{2}m^{2}\right] ^{1/2}.
\label{nup}
\end{equation}%
For the stability of the Poincar\'{e} vacuum the parameter $\nu _{+}$ should
be real \cite{Brei82,Brei82b,Mezi85}. For $\nu _{+}\geq 1$ from the
normalizability condition it follows that $c_{2}=0$ in the linear
combination of the cylinder functions. For $0\leq \nu _{+}<1$ the modes with
$c_{2}\neq 0$ are normalizable. In this case one of the coefficients in the
linear combination is determined from the normalization condition and the
second one is fixed by the boundary condition on the AdS boundary. The
general class of allowed boundary condition has been discussed in \cite%
{Ishi03,Ishi04}. Here we will consider the special case of Dirichlet
boundary condition for which $c_{2}=0$ and $f(z)=c_{1}z^{D/2}J_{\nu
_{+}}(\lambda z)$. The normalized mode functions are expressed as%
\begin{equation}
\varphi _{\sigma }^{(\pm )}(x)=\left( \frac{\pi ^{-p}a^{1-D}\lambda }{%
2^{p+1}\omega V_{q}}\right) ^{\frac{1}{2}}z^{\frac{D}{2}}e^{i\mathbf{k}_{p-1}%
\mathbf{x}_{p-1}+i\mathbf{k}_{q}\mathbf{x}_{q}\mp i\omega t}J_{\nu
_{+}}(\lambda z).  \label{phipm}
\end{equation}%
The modes are specified by the set $\sigma =(\lambda ,\mathbf{k}_{p-1},%
\mathbf{k}_{q})$ with $0\leq \lambda <\infty $ and the energy is given by $%
\omega =\sqrt{\lambda ^{2}+\mathbf{k}_{p-1}^{2}+\mathbf{k}_{q}^{2}}$.

With the mode functions (\ref{phipm}), the Hadamard function takes the form%
\begin{equation}
G(x,x^{\prime })=\frac{a^{1-D}(zz^{\prime })^{\frac{D}{2}}}{(2\pi
)^{p-1}V_{q}}\sum_{\mathbf{n}_{q}}\int d\mathbf{k}_{p-1}\,e^{i\mathbf{k}%
_{p-1}\cdot \Delta \mathbf{x}_{p-1}+i\mathbf{k}_{q}\cdot \Delta \mathbf{x}%
_{q}}\int_{0}^{\infty }d\lambda \,\frac{\lambda }{\omega }J_{\nu
_{+}}(\lambda z)J_{\nu _{+}}(\lambda z^{\prime })\cos (\omega \Delta t).
\label{GHAdS}
\end{equation}%
Similar to the cases of the locally Minkowski and dS geometries, we aplly to
the series over $n_{r}$ the summation formula (\ref{APs}) to see the
representation%
\begin{eqnarray}
G(x,x^{\prime }) &=&\frac{2(zz^{\prime })^{\frac{D}{2}}L_{r}}{(2\pi )^{\frac{%
p+1}{2}}V_{q}a^{D-1}}\sum_{\mathbf{n}_{q}}e^{i\mathbf{k}_{q-1}\cdot \Delta
\mathbf{x}_{q-1}+in_{r}\tilde{\alpha}_{r}}\int_{0}^{\infty }d\lambda
\,\lambda J_{\nu _{+}}(\lambda z)J_{\nu _{+}}(\lambda z^{\prime })  \notag \\
&&\times \left( \lambda ^{2}+\mathbf{k}_{q-1}^{2}\right) ^{\frac{p-1}{2}}f_{%
\frac{p-1}{2}}\left( \sqrt{\lambda ^{2}+\mathbf{k}_{q-1}^{2}}\sqrt{|\Delta
\mathbf{x}_{p-1}|^{2}+\left( \Delta x^{r}-n_{r}L_{r}\right) ^{2}-\left(
\Delta t\right) ^{2}}\right) .  \label{GHAdS1}
\end{eqnarray}%
The term with $n_{r}=0$ in this representation corresponds to the Hadamard
function in the geometry where the $r$th dimension is decompactified.

Another representation for the function (\ref{GHAdS}) is obtained in \cite%
{Beze15} by using an integral representation for the ratio $\cos (\omega
\Delta t)/\omega $. The integral over $\lambda $ is expressed in terms of
the modified Bessel function. Integrating over the components of the
momentum along uncompact dimensions and applying to the series the Poisson
resummation formula the Hadamard function is expressed as%
\begin{equation}
G(x,x^{\prime })=\frac{a^{1-D}}{(2\pi )^{D/2}}\sum_{\mathbf{n}_{q}}e^{i%
\tilde{\mathbf{\alpha }}\cdot \mathbf{n}_{q}}\int_{0}^{\infty
}dx\,x^{D/2-1}I_{\nu }(x)e^{-v_{\mathbf{n}_{q}}x},  \label{GHAdS2}
\end{equation}%
where%
\begin{equation}
v_{\mathbf{n}_{q}}=1+\frac{1}{2zz^{\prime }}\left[ (\Delta z)^{2}+(\Delta
\mathbf{x}_{p-1})^{2}+\sum_{i=p+1}^{D}\left( \Delta x^{i}-L_{i}n_{i}\right)
^{2}-(\Delta t)^{2}\right] ,  \label{vn}
\end{equation}%
and $\Delta z=z-z^{\prime }$. By using the result from \cite{PrudV2} for the
integral in (\ref{GHAdS2}), the following representation is obtained:
\begin{equation}
G(x,x^{\prime })=\frac{2a^{1-D}}{(2\pi )^{\frac{D+1}{2}}}\sum_{\mathbf{n}%
_{q}}e^{i\tilde{\mathbf{\alpha }}\cdot \mathbf{n}_{q}}q_{\nu _{+}-\frac{1}{2}%
}^{\frac{D-1}{2}}(v_{\mathbf{n}_{q}}),  \label{GHAdS3}
\end{equation}%
where the function $q_{\alpha }^{\mu }(x)$ is expressed in terms of the
associated Legendre function of the second kind, $Q_{\alpha }^{\mu }(x)$
(for the expression in terms of the hypergeometric function see Appendix \ref%
{sec:App1}):
\begin{equation}
q_{\nu -\frac{1}{2}}^{\mu }(x)=\frac{e^{-i\pi \mu }Q_{\nu -\frac{1}{2}}^{\mu
}(x)}{(x^{2}-1)^{\frac{\mu }{2}}}.  \label{qmu}
\end{equation}%
The contribution in (\ref{GHAdS3}) corresponding to the term $\mathbf{n}%
_{q}=0$ presents the Hadamard function in AdS spacetime with Poincar\'{e}
coordinates $-\infty <x^{\mu }<+\infty $ for $\mu =2,3,\ldots ,D$. The
divergences in the coincidence limit are contained in that part. The
topological contribution with $\mathbf{n}_{q}\neq 0$ is finite in that limit
and can be directly used in evaluating the current density.

As before, the charge density and the components of the current density
along uncompact dimensions vanish: $\left\langle j^{\mu }\right\rangle =0$
for $\mu =0,1,\ldots ,p$. Combining the formulas (\ref{jmuVev}) and (\ref%
{GHAdS1}), for the component along the $r$th dimension we find%
\begin{eqnarray}
\langle j^{r}\rangle &=&\frac{4ez^{D+2}L_{r}^{2}}{(2\pi )^{\frac{p+1}{2}%
}a^{D+1}V_{q}}\sum_{n_{r}=1}^{\infty }n_{r}\sin \left( n_{r}\tilde{\alpha}%
_{r}\right) \sum_{\mathbf{n}_{q-1}}\int_{0}^{\infty }d\lambda \,\lambda
J_{\nu _{+}}^{2}(\lambda z)  \notag \\
&&\times \left( \lambda ^{2}+\mathbf{k}_{q-1}^{2}\right) ^{\frac{p+1}{2}}f_{%
\frac{p+1}{2}}\left( n_{r}L_{r}\sqrt{\lambda ^{2}+\mathbf{k}_{q-1}^{2}}%
\right) .  \label{jrAdS}
\end{eqnarray}%
By applying the formula (\ref{RelSer}) with $s=p$ to the series over $%
\mathbf{n}_{q-1}$ one gets
\begin{equation}
\langle j^{r}\rangle =\frac{4ez^{D+2}L_{r}}{(2\pi )^{\frac{D}{2}}a^{D+1}}%
\sum_{n_{r}=1}^{\infty }n_{r}\sin \left( n_{r}\tilde{\alpha}_{r}\right)
\sum_{\mathbf{n}_{q-1}}\cos (\mathbf{n}_{q-1}\cdot \boldsymbol{\tilde{\alpha}%
}_{q-1})\int_{0}^{\infty }d\lambda \,\lambda ^{D+1}J_{\nu _{+}}^{2}(\lambda
z)f_{\frac{D}{2}}(\lambda g(\mathbf{L}_{q},\mathbf{n}_{q})).  \label{jrAdS1}
\end{equation}%
The integral in (\ref{jrAdS1}) is expressed in terms of the function (\ref%
{qmu}) \cite{Grad07} (note that there is a misprint in the similar integral
given in \cite{PrudV2}) and the current density is presented in the form
\cite{Beze15}
\begin{equation}
\langle j^{r}\rangle =\frac{4eL_{r}}{(2\pi )^{\frac{D+1}{2}}a^{D+1}}%
\sum_{n_{r}=1}^{\infty }n_{r}\sin (\tilde{\alpha}_{r}n_{r})\sum_{\mathbf{n}%
_{q-1}}\,\cos (\tilde{\mathbf{\alpha }}_{q-1}\cdot \mathbf{n}_{q-1})q_{\nu
_{+}-\frac{1}{2}}^{\frac{D+1}{2}}\left( 1+\frac{g^{2}(\mathbf{L}_{q},\mathbf{%
n}_{q})}{2z^{2}}\right) .  \label{jrAdS2}
\end{equation}%
This representation could be directly obtained by using the Hadamard
function in the form (\ref{GHAdS3}) and the relation $\partial _{x}q_{\alpha
}^{\mu }(x)=-q_{\alpha }^{\mu +1}(x)$ for the function (\ref{qmu}).

\section{Features of the Current Density}

\label{sec:Feat}

\subsection{General Features}

The physical component of the charge density is given by $\left\langle j_{%
\mathrm{(p)}}^{r}\right\rangle =\sqrt{|g_{rr}|}\left\langle
j^{r}\right\rangle $. It determines the charge flux through the spatial
hypersurface $x^{r}=\mathrm{const}$, expressed as $n_{r}\left\langle
j^{r}\right\rangle $, with $n_{r}=\sqrt{|g_{rr}|}$ being the corresponding
normal. The expressions obtained above for the $r$th component of the
current density can be combined as%
\begin{equation}
\left\langle j_{\mathrm{(p)}}^{r}\right\rangle =\frac{2eL_{\mathrm{(p)}r}}{%
\left( 2\pi \right) ^{\frac{D+1}{2}}a^{D+1}}\sum_{\mathbf{n}_{q}}n_{r}\sin
\left( \mathbf{n}_{q}\cdot \mathbf{\tilde{\alpha}}_{q}\right) F_{D}(am,g(%
\mathbf{L}_{\mathrm{(p)}q}/a,\mathbf{n}_{q})),  \label{jrc}
\end{equation}%
where we have defined the function%
\begin{eqnarray}
F_{D}(am,x) &=&\left( am\right) ^{D+1}f_{\frac{D+1}{2}}(amx),\mathrm{%
\;for\;LM},  \notag \\
F_{D}(am,x) &=&\sqrt{\frac{2}{\pi }}\int_{0}^{\infty }du\,u^{D+1}Z(u)f_{%
\frac{D}{2}}(ux),\mathrm{\;for\;LdS,LAdS},  \label{FD}
\end{eqnarray}%
with%
\begin{eqnarray}
Z(u) &=&\left[ I_{-\nu }(u)+I_{\nu }(u)\right] K_{\nu }(u),\mathrm{\;for\;LdS%
},  \notag \\
Z(u) &=&\pi J_{\nu _{+}}^{2}(u),\mathrm{\;for\;LAdS}.  \label{Znu}
\end{eqnarray}%
Alternative expressions for locally dS and AdS geometries are obtained from (%
\ref{jdS3}) and (\ref{jrAdS2}):%
\begin{eqnarray}
F_{D}(am,x) &=&\frac{1}{2}p_{\nu -\frac{1}{2}}^{-\frac{D+1}{2}}\left(
x^{2}/2-1\right) ,\mathrm{\;for\;LdS},  \notag \\
F_{D}(am,x) &=&q_{\nu _{+}-\frac{1}{2}}^{\frac{D+1}{2}}\left(
x^{2}/2+1\right) ,\mathrm{\;for\;LAdS}.  \label{FD2}
\end{eqnarray}%
For even values of the spatial dimension $D$ the functions (\ref{FD2}) are
expressed in terms of elementary functions. The corresponding
representations are given by the formulas (\ref{FDadsEv}) and (\ref{FDdSEv}%
). In odd number of spatial dimensions the expressions for the functions (%
\ref{FD2}) in terms of the Legendre functions $P_{\nu -\frac{1}{2}}(u)$ and $%
Q_{\nu _{+}-\frac{1}{2}}(u)$ are given by (\ref{FDod}).

In asymptotic analysis for some limiting cases it is more convenient to use
the representations (\ref{jr}), (\ref{jdS}), and (\ref{jrAdS}). For LdS and
LAdS geometries the corresponding formulas can be combined by using (\ref%
{Znu}):%
\begin{eqnarray}
\left\langle j_{\mathrm{(p)}}^{r}\right\rangle &=&\frac{8ea^{-p-2}L_{\mathrm{%
(p)}r}^{2}}{\left( 2\pi \right) ^{\frac{p+3}{2}}V_{q}^{\mathrm{(p)}}}%
\sum_{n_{r}=1}^{\infty }n_{r}\sin (n_{r}\tilde{\alpha}_{r})\sum_{\mathbf{n}%
_{q-1}}\int_{0}^{\infty }du\,uZ(u)  \notag \\
&&\times \left( u^{2}+a^{2}\mathbf{k}_{q-1}^{\mathrm{(p)}2}\right) ^{\frac{%
p+1}{2}}f_{\frac{p+1}{2}}(n_{r}L_{\mathrm{(p)}r}/a\sqrt{u^{2}+a^{2}\mathbf{k}%
_{q-1}^{\mathrm{(p)}2}}).  \label{jrc2}
\end{eqnarray}%
with the notation%
\begin{equation}
\mathbf{k}_{q-1}^{\mathrm{(p)}2}=\sum_{i=p+1,\neq r}^{D}\left( \frac{2\pi
n_{i}+\tilde{\alpha}_{i}}{L_{\mathrm{(p)}i}}\right) ^{2}.  \label{kgp}
\end{equation}%
Note that the vector $\mathbf{k}_{q-1}^{\mathrm{(p)}}$ is the physical
momentum in the compact subspace with the set of coordinates $%
(x^{p+1},\ldots ,x^{r-1},x^{r+1},\ldots ,x^{D})$.

First of all, we see that the current density along the $r$th dimension is
an even periodic function of the parameters $\tilde{\alpha}_{i}$, $i\neq r$,
with the period $2\pi $ and an odd periodic function of $\tilde{\alpha}_{r}$
with the same period. This corresponds to the periodicity with respect to
the magnetic flux with the period equal to the flux quantum. From the
formula (\ref{jrc}) it follows that the physical component $%
n_{r}\left\langle j^{r}\right\rangle $ depends on the lengths of compact
dimensions and on the coordinates through the ratios $L_{\mathrm{(p)}i}/a$, $%
i=p+1,\ldots ,D$. They present the proper lengths of compact dimensions
measured in units of the curvature radius. This feature is related to the
maximal symmetry of the dS and AdS spacetimes.

\begin{figure}[tbph]
\begin{center}
\epsfig{figure=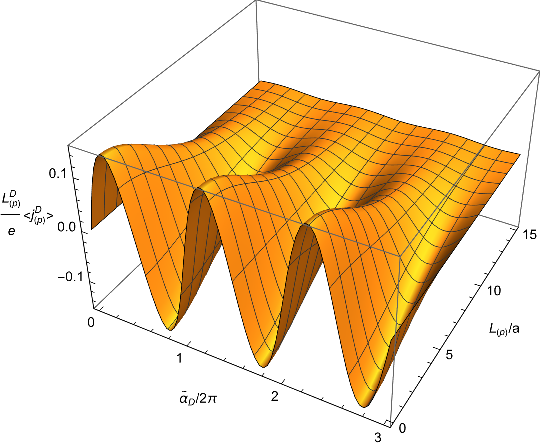,width=7.5cm,height=6cm}
\end{center}
\caption{The current density in the $D=4$ LM\ spacetime versus the parameter
$\tilde{\protect\alpha}_{D}/(2\protect\pi )$ and the proper length of the
compact dimension (in units of $a$). The graph is plotted for $ma=0.5$.}
\label{fig3DM}
\end{figure}

The numerical examples below will be given for models with a single compact
dimension $x^{D}$ having the length $L=L_{D}$. In Figure \ref{fig3DM} we
present the dependence of the respective current density, multiplied by $L_{%
\mathrm{(p)}}^{D}/e$, as a function of the parameter $\tilde{\alpha}%
_{D}/(2\pi )$ and of the proper length of the compact dimension $L_{\mathrm{%
(p)}}=L_{\mathrm{(p)}D}$ in the LM spacetime with $D=4$. In the numerical
evaluation we have taken $ma=0.5$. In the LM bulk the current density does
not depend on the curvature coupling parameter and $L_{\mathrm{(p)}}=L$. As
it follows from (\ref{jrp1m0}), for a massless field the dimensionless
combination $L_{\mathrm{(p)}}^{D}\left\langle j_{\mathrm{(p)}%
}^{D}\right\rangle /e$ does not depend on $L_{\mathrm{(p)}}$.

\begin{figure}[tbph]
\begin{center}
\begin{tabular}{cc}
\epsfig{figure=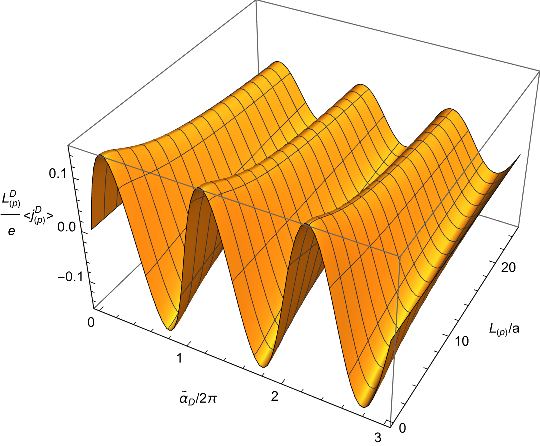,width=7.5cm,height=6cm} & \quad %
\epsfig{figure=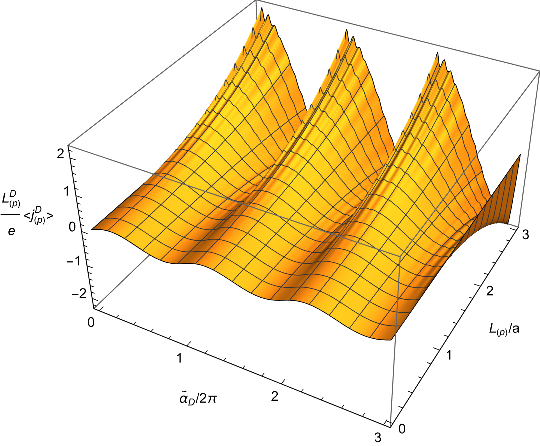,width=7.5cm,height=6cm}%
\end{tabular}%
\end{center}
\caption{The current density in the $D=4$ LdS spacetime, multiplied by $L_{%
\mathrm{(p)}}^{D}/e$, versus the parameter $\tilde{\protect\alpha}_{D}/(2%
\protect\pi )$ and the proper length of the compact dimension for $ma=0.5$.
The left and right panels correspond to conformally and minimally coupled
fields, respectively. }
\label{figdS3D}
\end{figure}

\begin{figure}[tbph]
\begin{center}
\begin{tabular}{cc}
\epsfig{figure=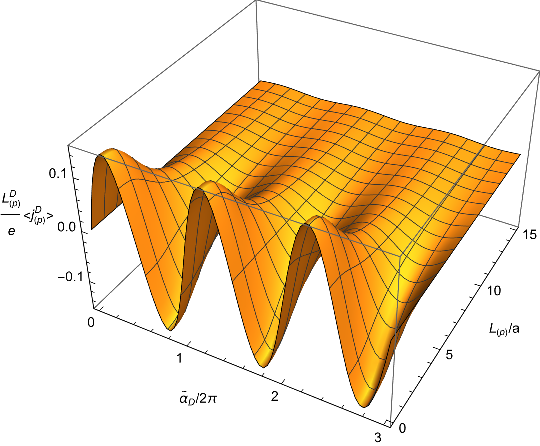,width=7.5cm,height=6cm} & \quad %
\epsfig{figure=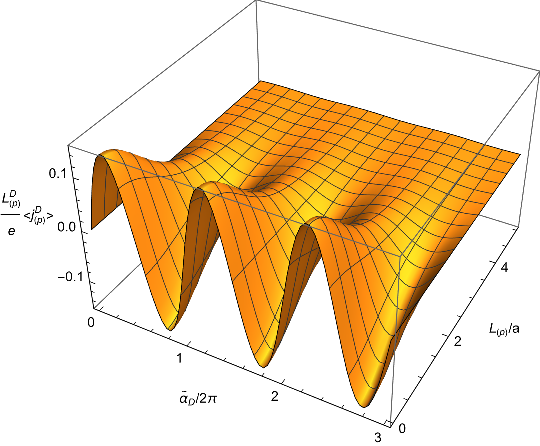,width=7.5cm,height=6cm}%
\end{tabular}%
\end{center}
\caption{The same as in Figure \protect\ref{figdS3D} for the LAdS bulk.}
\label{figAdS3D}
\end{figure}

The current densities for the $D=4$ LdS and LAdS background geometries and
for $ma=0.5$ are plotted in Figures \ref{figdS3D} and \ref{figAdS3D}. The
left and right panels on both figures correspond to conformally and
minimally coupled fields.

\subsection{Conformal Coupling and Minkowskian Limit}

Let us consider special cases of general formulas. For a conformally coupled
massless field one has $\xi =\xi _{D}$ and $\nu =\nu _{+}=1/2$. The current
density for the Minkowskian case does not depend on the curvature coupling
parameter and from (\ref{jr3}) we get%
\begin{equation}
\left\langle j^{r}\right\rangle _{\mathrm{LM}}=eL_{r}\frac{\Gamma \left(
\frac{D+1}{2}\right) }{\pi ^{\frac{D+1}{2}}}\sum_{\mathbf{n}_{q}}\frac{%
n_{r}\sin \left( \mathbf{n}_{q}\cdot \mathbf{\tilde{\alpha}}_{q}\right) }{%
g^{D+1}(\mathbf{L}_{q},\mathbf{n}_{q})}.  \label{jrMm0}
\end{equation}%
The corresponding functions $F_{D}(am,x)$ for dS and AdS geometries are
obtained from (\ref{FD}) by taking into account that $\left[ I_{-\nu
}(u)+I_{\nu }(u)\right] K_{\nu }(u)=1/u$ for $\nu =1/2$ and $J_{\nu _{+}}(u)=%
\sqrt{2/\pi u}\sin u$ for $\nu _{+}=1/2$. The integrals are evaluated by
using the formulas from \cite{PrudV2} and we get%
\begin{eqnarray}
\left\langle j^{r}\right\rangle _{\mathrm{LdS}} &=&\left( \frac{\eta }{a}%
\right) ^{D+1}\left\langle j^{r}\right\rangle _{\mathrm{LM}},  \notag \\
\left\langle j^{r}\right\rangle _{\mathrm{LAdS}} &=&\left( \frac{z}{a}%
\right) ^{D+1}\left\langle j^{r}\right\rangle _{\mathrm{LM}}^{(1)},
\label{jrdSm0}
\end{eqnarray}%
where
\begin{equation}
\left\langle j^{r}\right\rangle _{\mathrm{LM}}^{(1)}=eL_{r}\frac{\Gamma
\left( \frac{D+1}{2}\right) }{\pi ^{\frac{D+1}{2}}}\sum_{\mathbf{n}%
_{q}}n_{r}\sin \left( \mathbf{n}_{q}\cdot \mathbf{\tilde{\alpha}}_{q}\right) %
\left[ \frac{1}{g^{D+1}(\mathbf{L}_{q},\mathbf{n}_{q})}-\frac{1}{\left(
4z^{2}+g^{2}(\mathbf{L}_{q},\mathbf{n}_{q})\right) ^{\frac{D+1}{2}}}\right] .
\label{jrMm0b}
\end{equation}%
The result (\ref{jrdSm0}) for the dS bulk was expected from the conformal
relation between the problems in the Minkowski and dS geometries with the
same range of the spatial coordinates and between the Bunch-Davies and
Minkowski vacua. For the AdS bulk the contribution of the first term in the
square brackets of (\ref{jrMm0b}) will give the Minkowskian current density
multiplied by the conformal factor $(z/a)^{D+1}$. The presence of the part
coming from the second term in the square brackets is related to the
boundary condition on the AdS boundary at $z=0$. Because of that condition
the problem on the AdS bulk for a conformally coupled massless field is
conformally related to the corresponding problem in the Minkowski bulk with
an additional boundary at $z=0$ with Dirichlet boundary condition for the
field. The VEV (\ref{jrMm0b}) is the current density for a conformally
coupled massless field in the region $0<z<\infty $ of the locally Minkowski
bulk with Dirichlet boundary at $z=0$.

\begin{figure}[tbph]
\begin{center}
\begin{tabular}{cc}
\epsfig{figure=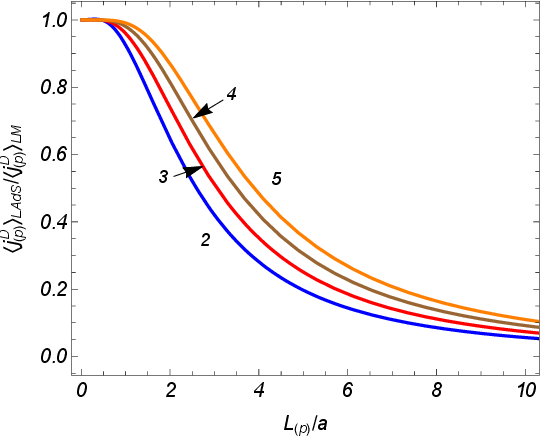,width=7.5cm,height=6cm} & \quad %
\epsfig{figure=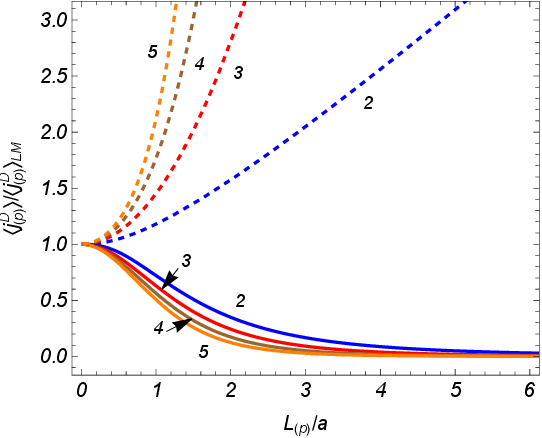,width=7.5cm,height=6cm}%
\end{tabular}%
\end{center}
\caption{The left panel presents the ratio of the current densities for a
conformally coupled massless scalar field in LAdS and LM spacetimes, with a
single compact dimension of the proper length $L_{\mathrm{(p)}}=L_{\mathrm{%
(p)}D}$, versus the ratio $L_{\mathrm{(p)}}/a$. On the right panel the ratio
$\left\langle j^{D}\right\rangle /\left\langle j^{D}\right\rangle _{\mathrm{%
LM}}$ is plotted for a minimally coupled massless scalar field in LAdS (full
curves, $\left\langle j^{D}\right\rangle =\left\langle j^{D}\right\rangle _{%
\mathrm{LAdS}}$) and LdS (dashed curves, $\left\langle j^{D}\right\rangle
=\left\langle j^{D}\right\rangle _{\mathrm{LdS}}$). The numbers bear the
curves present the corresponding spatial dimension. The different behaviour
for LAdS and LdS geometries in the region of large compact dimensions 
will be clarified below by the asymptotic analysis.}
\label{figm0}
\end{figure}

In Figure \ref{figm0}, for a massless field, we have plotted the dependence
of the ratios $\left\langle j^{D}\right\rangle _{\mathrm{LAdS}}/\left\langle
j^{D}\right\rangle _{\mathrm{LM}}$ and $\left\langle j^{D}\right\rangle _{%
\mathrm{LdS}}/\left\langle j^{D}\right\rangle _{\mathrm{LM}}$ on the proper
length $L_{\mathrm{(p)}}=L_{\mathrm{(p)}D}$ of a single compact dimension
(in units of the curvature radius $a$). It is assumed that the compact
dimension has the same proper length in LAdS, LdS and LM spacetimes. The
left and right panels correspond to conformally and minimally coupled
fields, respectively, and the numbers near the curves present the values of
the respective spatial dimension. For a conformally coupled field $%
\left\langle j^{D}\right\rangle _{\mathrm{LdS}}/\left\langle
j^{D}\right\rangle _{\mathrm{LM}}=1$ and only the case of the LAdS bulk is
depicted on the left panel. The full and dashed curves on the right panel
correspond to the LAdS and LdS geometries, respectively. As seen from the
graphs, for massless fields the decay of the current density, as a function
of the proper length of the compact dimension, is stronger in the LAdS
spacetime (compared to the case of the LM bulk). For a minimally coupled
field in the LdS geometry the fall-off of the current density is stronger in
the LM spacetime. For small values of the proper length compared to the
curvature radius the effect of the gravitational field is weak and the ratio
$\left\langle j^{D}\right\rangle /\left\langle j^{D}\right\rangle _{\mathrm{%
LM}}$ tends to 1. All these features will be confirmed below by asymptotic
analysis.

Now let us check the Minkowskian limit for dS and AdS geometries. As seen
from (\ref{dsdS}) and (\ref{dsAdS}), it is obtained taking $a\rightarrow
\infty $ for fixed spacetime coordinates $(t,x^{1},\ldots ,x^{D})$ and $%
(t,y,x^{2},\ldots ,x^{D})$ for the dS and AdS cases, respectively. For large
curvature radius one has $\nu \approx iam$, $\eta \approx a-t$ in LdS bulk
and $\nu _{+}\approx am$, $z\approx a+y$ for LAdS. In the case of LAdS we
need the asymptotic expression of the function $q_{\nu
_{+}-1/2}^{(D+1)/2}\left( u^{2}/2+1\right) $ for $\nu _{+}\gg 1$ and $u\ll 1$%
. That expression is obtained by using the uniform asymptotic expression of
the associated Legendre function of the second kind for large degree and
fixed order, given in \cite{Olve10}. With that asymptotic, it can be checked
that in the limit under consideration%
\begin{equation}
q_{\nu _{+}-1/2}^{\frac{D+1}{2}}\left( 1+\frac{g^{2}(\mathbf{L}_{\mathrm{(p)}%
q},\mathbf{n}_{q})}{2a^{2}}\right) \approx \left( am\right) ^{D+1}f_{\frac{%
D+1}{2}}\left( mg(\mathbf{L}_{q},\mathbf{n}_{q})\right) ,  \label{LimMink}
\end{equation}%
confirming the transition to the Minkowskian result. In the case of LdS it
is convenient to use the representation (\ref{FD}) for the function $%
F_{D}(am,x)$. In the limit at hand $\nu \approx iam$, $am\gg 1$. The
corresponding uniform asymptotic expansions for the functions $I_{\pm \nu
}(u)$ and $K_{\nu }(u)$ can be found, for example, in ~\cite{Duns90,Milt09}.
From those expansions it can be seen that the dominant contribution to the
integral in the expression for $F_{D}(am,x)$ comes from the region $u>am$,
where%
\begin{equation}
\left[ I_{\nu }(u)+I_{-\nu }(u)\right] K_{\nu }(u)\sim \frac{1}{\sqrt{%
u^{2}-a^{2}m^{2}}}.  \label{IKas}
\end{equation}%
with $\nu \approx iam$. The respective integral is evaluated by using the
formula%
\begin{equation}
\int_{0}^{\infty }du\,\left( u^{2}+b^{2}\right) ^{\mu }f_{\mu }(c\sqrt{%
u^{2}+b^{2}})=\sqrt{\frac{\pi }{2}}b^{2\mu +1}f_{\mu +\frac{1}{2}}(cb),
\label{Int1}
\end{equation}%
and we see that to the leading order%
\begin{equation}
F_{D}(am,x)\approx \left( ma\right) ^{D+1}f_{\frac{D+1}{2}}(max),
\label{LimMink2}
\end{equation}%
which coincides with the Minkowskian result.

\subsection{Large and Small Proper Lengths of Compact Dimensions}

For small values of the proper length of the $r$th compact dimension, $L_{%
\mathrm{(p)}r}\ll a,1/m$, we first consider the contribution in (\ref{jrc})
of the terms for which at least one of $n_{i}$, $i\neq r$, is different from
zero. For that part the dominant contribution to the series over $n_{r}$
comes from the terms with large values $|n_{r}|$ and we replace the
corresponding summation by the integration. The corresponding integral
involving the product of the sin and Macdonald functions is evaluated by
using the formula from \cite{Grad07} and is expressed in terms of the
Macdonald function with large argument. By using the corresponding
asymptotic we see that the contribution of the term for which at least one
of $n_{i}$, $i\neq r$, is not zero is suppressed by the factor $\exp [-g(%
\mathbf{L}_{\mathrm{(p)}q-1}/a,\mathbf{n}_{q-1})\tilde{\alpha}_{r}a/L_{%
\mathrm{(p)}r}]$, where
\begin{equation}
g^{2}(\mathbf{L}_{\mathrm{(p)}q-1}/a,\mathbf{n}_{q-1})=\sum_{i=p+1,\neq
r}^{D}n_{i}^{2}\frac{L_{i}^{2}}{a^{2}}.  \label{gqm1}
\end{equation}%
For the contribution of the terms with $n_{i}=0$, $i\neq r$, in (\ref{jrc})
the argument $x$ of the function $F_{D}(am,x)$ is small. By using the
corresponding asymptotic for the Macdonald function we see that in the case
of LM bulk%
\begin{equation}
F_{D}(am,x)\approx \frac{2^{\frac{D-1}{2}}}{x^{D+1}}\Gamma \left( \frac{D+1}{%
2}\right) ,\;x\ll 1.  \label{FDsmx}
\end{equation}%
In the cases of the LdS and LAdS geometries we note that the main
contribution to the integrals in (\ref{FD}) comes from the region with large
values of $u$. By using the respective approximations for the functions $%
\left[ I_{-\nu }(u)+I_{\nu }(u)\right] K_{\nu }(u)$ and $J_{\nu _{+}}^{2}(u)$
and evaluating the integrals, we can see that the corresponding asymptotics
are given by the same expression (\ref{FDsmx}). Hence, in the limit $L_{%
\mathrm{(p)}r}\ll a,1/m$ the dominant contribution to the current density
comes from the modes with $n_{i}=0$, $i\neq r$, and to the leading order%
\begin{equation}
\left\langle j_{\mathrm{(p)}}^{r}\right\rangle \approx \frac{2e\Gamma \left(
\frac{D+1}{2}\right) }{\pi ^{\frac{D+1}{2}}L_{\mathrm{(p)}r}^{D}}%
\sum_{n_{r}=1}^{\infty }\frac{\sin \left( n_{r}\tilde{\alpha}_{r}\right) }{%
n_{r}^{D}}.  \label{jrpsm}
\end{equation}%
The expression in the right-hand side presents the current density for a
massless scalar field in the LM spacetime with spatial topology $%
R^{D-1}\times S^{1}$ with a single compact dimension $x^{r}$ with length $%
L_{r}=L_{\mathrm{(p)}r}$. For small values of $L_{\mathrm{(p)}r}$ the
dominant contribution to the VEVs comes from the vacuum fluctuations with
small values of the wavelength (compared to the curvature radius) and the
effects of gravity are weak.

It is expected that the effects of gravity on the vacuum currents will be
essential for proper lengths of compact dimensions of the order or larger
than the curvature radius. We start the consideration for large values of
the lengths with the LM case, assuming that $L_{r}$ is much larger than the
other length scales of the model. From the formula (\ref{jr}) it follows
that the dominant contribution comes from the modes with $n_{i}=0$, $i\neq r$%
, for which $\omega _{\mathbf{n}_{q-1}}=\omega _{0r}=\sqrt{k_{\mathbf{n}%
_{q-1}}^{(0)2}+m^{2}}$ with
\begin{equation}
k_{\mathbf{n}_{q-1}}^{(0)2}=\sum\nolimits_{l=p+1,\neq r}^{D}\tilde{\alpha}%
_{l}^{2}/L_{l}^{2}.  \label{k0r}
\end{equation}%
The beahvior of the current density is essentially different depending
whether $\omega _{0r}$ is zero or not. In the first case the leading term in
the current density is given as%
\begin{equation}
\left\langle j^{r}\right\rangle _{\mathrm{LM}}\approx \frac{2e\Gamma (\frac{p%
}{2}+1)}{\pi ^{\frac{p}{2}+1}L_{r}^{p}V_{q}}\sum_{n_{r}=1}^{\infty }\frac{%
\sin (n_{r}\tilde{\alpha}_{r})}{n_{r}^{p+1}}.  \label{jrMlLm0}
\end{equation}%
Comparing with (\ref{jrpsm}) we see that the right-hand side of (\ref%
{jrMlLm0}), multiplied by $V_{q-1}=V_{q}/L_{r}$ presents the current density
for a massless scalar field in $(p+2)$-dimensional LM spacetime with spatial
topology $R^{p}\times S^{1}$ having a single compact dimension $x^{r}$. For $%
\omega _{0r}\neq 0$ the dominant contribution to the current density is
induced by the mode with $n_{r}=1$ and, to the leading order,%
\begin{equation}
\left\langle j^{r}\right\rangle _{\mathrm{LM}}\approx \frac{2e\sin (\tilde{%
\alpha}_{r})\omega _{0r}^{\frac{p+1}{2}}}{(2\pi )^{\frac{p+1}{2}}L_{r}^{%
\frac{p-1}{2}}V_{q}}e^{-L_{r}\omega _{0r}}.  \label{jrT0large}
\end{equation}%
In particular, for the model with a single compact dimension $x^{D}$ one has
$p=D-1$ and the asymptotic (\ref{jrT0large}) takes the form%
\begin{equation}
\left\langle j^{D}\right\rangle _{\mathrm{LM}}\approx \frac{2e\sin (\tilde{%
\alpha}_{D})m^{\frac{D}{2}}}{(2\pi L_{D})^{\frac{D}{2}}}e^{-mL_{D}},
\label{jDMlarge1}
\end{equation}%
where $mL_{D}\gg 1$.

For LdS and LAdS geometries and for large values of the proper length $L_{%
\mathrm{(p)}r}$ it is more convenient to use the representations (\ref{jdS})
and (\ref{jrAdS}). The dominant contribution comes from the term in the
summation with $\mathbf{n}_{q-1}=0$ ($n_{l}=0$ for $l\neq r$) and from the
integration region near the lower limit. Two cases should be considered
separately. The first one corresponds to the phases $\tilde{\alpha}_{i}=0$, $%
i\neq r$. With these values and for LAdS bulk and for LdS bulk in the case $%
\nu >0$ the leading order term is expressed as%
\begin{equation}
\langle j_{\mathrm{(p)}}^{r}\rangle \approx \frac{4ea^{1+2\mu }B_{D}(am)}{%
\pi ^{\frac{p+1}{2}}V_{q}^{\mathrm{(p)}}L_{\mathrm{(p)}r}^{p+1+2\mu }}\Gamma
\left( \frac{p+3}{2}+\mu \right) \sum_{n_{r}=1}^{\infty }\frac{\sin \left(
n_{r}\tilde{\alpha}_{r}\right) }{n_{r}^{p+2+2\mu }},  \label{jrLargeLp}
\end{equation}%
where%
\begin{equation}
\mu =\left\{
\begin{array}{cc}
-\nu , & \mathrm{for\;LdS} \\
\nu _{+}, & \mathrm{for\;LAdS}%
\end{array}%
\right. ,  \label{mudS}
\end{equation}%
and
\begin{equation}
B_{D}(am)=\left\{
\begin{array}{cc}
\Gamma \left( \nu \right) /(2\pi ), & \mathrm{for\;LdS} \\
1/\Gamma \left( \nu _{+}+1\right) , & \mathrm{for\;LAdS}%
\end{array}%
\right. .  \label{BD}
\end{equation}%
For LdS geometry and for imaginary values of $\nu $, $\nu =i|\nu |$, by
similar calculations the leading term is presented as%
\begin{equation}
\left\langle j_{\mathrm{(p)}}^{r}\right\rangle \approx \frac{4eC(ma)\eta
^{D+1}}{\pi ^{\frac{p+3}{2}}V_{q}a^{D}L_{r}^{p+1}}\sum_{n_{r}=1}^{\infty }%
\frac{\sin (n_{r}\tilde{\alpha}_{r})}{n_{r}^{p+2}}\cos \left[ 2|\nu |\ln
\left( n_{r}L_{r}/\eta \right) +\phi _{0}\right] ,  \label{jrLargLpim}
\end{equation}%
where the coefficient $C(ma)>0$ and the phase $\phi _{0}$ are defined by the
relation%
\begin{equation}
\Gamma (i|\nu |)\Gamma \left( \frac{p+3}{2}-i|\nu |\right) =C(ma)e^{i\phi
_{0}}.  \label{Cma}
\end{equation}%
In this case the current density exhibits an oscillatory behavior with the
amplitude decaying as $1/L_{\mathrm{(p)}r}^{p+1}$. Comparing (\ref{jrLargeLp}%
) and (\ref{jrLargLpim}) with (\ref{jrT0large}) we see that the
gravitational field essentially modifies the asymptotic behavior of the
current density for large values of the proper length $L_{\mathrm{(p)}r}$:
one has a power law decay in LdS and LAdS geometries instead of exponential
suppression for the LM bulk.

\begin{figure}[tbph]
\begin{center}
\begin{tabular}{cc}
\epsfig{figure=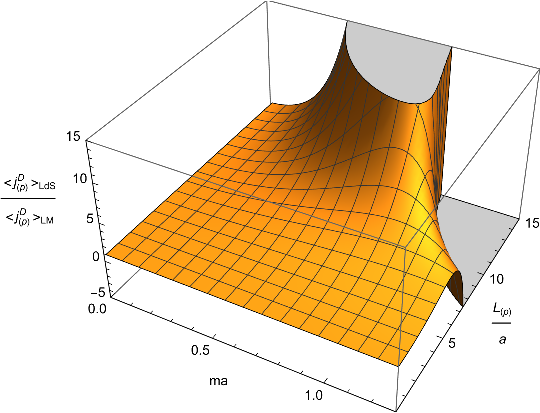,width=7.5cm,height=6cm} & \quad %
\epsfig{figure=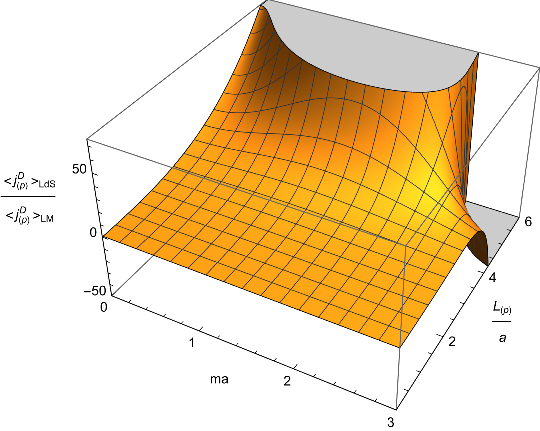,width=7.5cm,height=6cm}%
\end{tabular}%
\end{center}
\caption{The ratio of the current densities in the $D=4$ LdS and LM
spacetimes with the same proper lengths of the single compact dimension
versus the mass and the proper length (in units of $a$). The left and right
panels correspond to conformally and minimally coupled fields and the graphs
are plotted for $\tilde{\protect\alpha}_{D}=2\protect\pi /5$. }
\label{figreldS}
\end{figure}

In particular, the formulas (\ref{jrLargeLp}) and (\ref{jrLargLpim}) with $%
p=D-1$ describe the behavior of the current density for large values of $L_{%
\mathrm{(p)}D}$ in models with a single compact dimension $x^{D}$. In that
special case the asymptotic of the Minkowskian current density for a massive
field is described by (\ref{jDMlarge1}). In order to display the the
essential difference of the large $L_{\mathrm{(p)}r}$ asymptotics for LdS
and LAdS from that in the LM geometry, in Figures \ref{figreldS} and \ref%
{figrelAdS} we present the ratios $\left\langle j^{D}\right\rangle
/\left\langle j^{D}\right\rangle _{\mathrm{LM}}$ for $D=4$ LdS and LAdS
spacetimes, with a single compact dimension of the proper length $L_{\mathrm{%
(p)}}=L_{\mathrm{(p)}D}$, as functions of $ma$ and $L_{\mathrm{(p)}}/a$ for
fixed $\tilde{\alpha}_{D}=2\pi /5$. The ratios are evaluated for the same
values of the proper lengths in the LM, LdS and LAdS spacetimes and all the
quantities are measured in units of $a$.

\begin{figure}[tbph]
\begin{center}
\begin{tabular}{cc}
\epsfig{figure=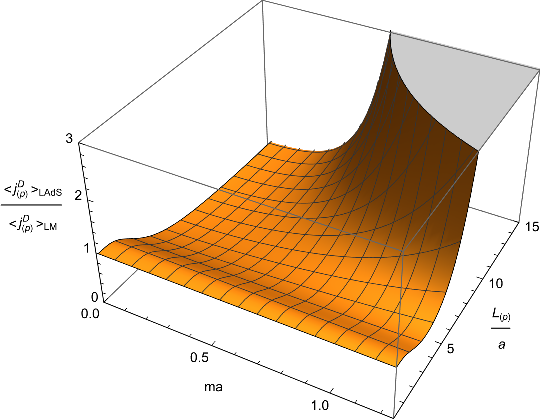,width=7.5cm,height=6cm} & \quad %
\epsfig{figure=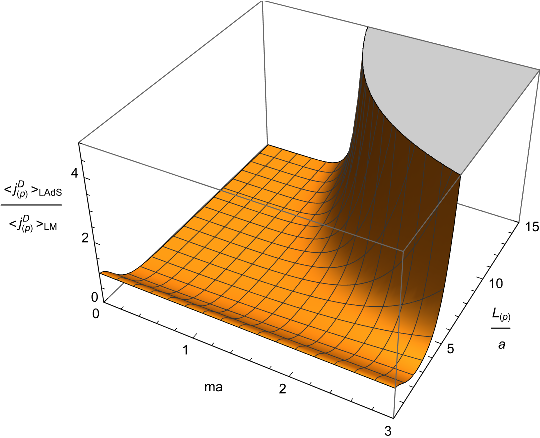,width=7.5cm,height=6cm}%
\end{tabular}%
\end{center}
\caption{The same as in Figure \protect\ref{figreldS} for the LAdS
spacetime. }
\label{figrelAdS}
\end{figure}

If at least one of the phases $\tilde{\alpha}_{i}$, $i\neq r$, is different
from zero and the proper length $L_{\mathrm{(p)}r}$ is large, we use the
asymptotic expression of the Macdonald function for large argument. For the
LAdS geometry and LdS geometry with positive values of $\nu $ the leading
contribution to the series over $n_{r}$ comes from the term $n_{r}=1$ and we
get%
\begin{equation}
\langle j_{\mathrm{(p)}}^{r}\rangle \approx \frac{2ea^{1+2\mu }B_{D}(am)\sin
\left( \tilde{\alpha}_{r}\right) }{2^{\frac{p}{2}+\mu }\pi ^{\frac{p}{2}%
}V_{q}^{\mathrm{(p)}}L_{\mathrm{(p)}r}^{p+1+2\mu }}(L_{r}|\mathbf{k}%
_{q-1}^{(0)}|)^{\frac{p}{2}+1+\mu }e^{-L_{r}|\mathbf{k}_{q-1}^{(0)}|},
\label{jrpLargeLp2}
\end{equation}%
where, as before, $\mu =-\nu $ and $\mu =\nu _{+}$ for LdS and LAdS. In the
case of LdS bulk and imaginary $\nu $ the leading order term takes the form%
\begin{equation}
\left\langle j_{\mathrm{(p)}}^{r}\right\rangle \approx \frac{4e\eta
^{D+1}\sin (\tilde{\alpha}_{r})}{\left( 2\pi \right) ^{\frac{p+1}{2}%
}a^{D}V_{q}L_{r}^{p+1}}\frac{(L_{r}|\mathbf{k}_{q-1}^{(0)}|)^{\frac{p}{2}%
+1}e^{-|\mathbf{k}_{q-1}^{(0)}|L_{r}}}{\sqrt{2|\nu |\sinh \left( \pi |\nu
|\right) }}\cos \left[ |\nu |\ln \left( |\mathbf{k}_{q-1}^{(0)}|\frac{L_{r}}{%
2}\right) -\mathrm{arg}(\Gamma (i|\nu |))\right] .  \label{jrLargLpim2}
\end{equation}%
In the case of LdS the different asymptotic behavior for positive and
imaginary values of $\nu $ is related to different asymptotics of the
function $\left[ I_{\nu }(x)+I_{-\nu }(x)\right] K_{\nu }(x)$ for small
arguments. For the LdS geometry and $\nu =0$ for the leading contribution we
get%
\begin{equation}
\left\langle j_{\mathrm{(p)}}^{r}\right\rangle \approx \frac{4e\sin (\tilde{%
\alpha}_{r})\eta ^{D+1}}{\left( 2\pi \right) ^{\frac{p}{2}%
+1}a^{D}V_{q}L_{r}^{p+1}}(|\mathbf{k}_{q-1}^{(0)}|L_{r})^{\frac{p}{2}+1}e^{-|%
\mathbf{k}_{q-1}^{(0)}|L_{r}}\ln \left( |\mathbf{k}_{q-1}^{(0)}|L_{r}\right)
.  \label{jrLargLpnu0}
\end{equation}

Now let us consider the asymptotics with respect to the length of the $l$th
dimension with $l\neq r$. For large values $L_{l}$ compared with the other
length scales and for $L_{\mathrm{(p)}l}/a\gg 1$ the leading contribution to
(\ref{jrc}) comes from the term with $n_{l}=0$. As expected, this leading
term coincides with the current density in the geometry where the $l$th
dimension is decompactified. The corrections induced by the respective
compactification are suppressed by the factor $e^{-mL_{l}}$ ($1/L_{l}^{D+1}$
for a massless field) in the LM bulk and by the factor $1/L_{l}^{D+2+2\mu }$
for LdS and LAdS geometries, where $\mu $ is given by (\ref{mudS}).

In the opposite limit of small values of $L_{l}$ it is more convenient to
use the representations (\ref{jr}) and (\ref{jrc2}). The behavior of the
current density is essentially different for the cases $\tilde{\alpha}_{l}=0$
and $\tilde{\alpha}_{l}\neq 0$. In the first case the dominant contribution
to the summation over $\mathbf{n}_{q-1}$ comes from the modes with $n_{l}=0$%
. For the LM bulk the leading term in the expansion of $L_{\mathrm{(p)}%
l}\left\langle j_{\mathrm{(p)}}^{r}\right\rangle $ coincides with the
current density in $D$-dimensional LM spacetime which is obtained from the
intital $(D+1)$-dimensional spacetime excluding the $l$th dimension. The
same is the case for LdS and LAdS bulks with the difference that in the
leading terms of the expansion for $L_{\mathrm{(p)}l}\left\langle j_{\mathrm{%
(p)}}^{r}\right\rangle $ the parameters $\nu $ and $\nu _{+}$ are defined
for $(D+1)$-dimensional spacetime whereas in the formula for $D$-dimensional
current density $\left\langle j_{\mathrm{(p)}}^{r}\right\rangle $ the
corresponding expressions for $\nu $ and $\nu _{+}$ are obtained from (\ref%
{nju}) and (\ref{nup}) by the replacement $D\rightarrow D-1$. For small
values of $L_{l}$ and $\tilde{\alpha}_{l}\neq 0$ the contribution of the
modes with $n_{r}=1$ and $n_{l}=0$ dominates in (\ref{jr}) and (\ref{jrc2}).
In the remaining summations over $n_{i}$, $i\neq r,l$, the main contribution
comes from large values $n_{i}$ and we replace the corresponding series by
integrations. In this way it can be seen that the current density $%
\left\langle j_{\mathrm{(p)}}^{r}\right\rangle $ is suppressed by the factor
$\exp (-L_{r}|\tilde{\alpha}_{l}|/L_{l})$.

\subsection{Fermionic currents}

In the discussion above we have considered the current densities for a
charged scalar field. Similar investigations for the massive Dirac field $%
\psi (x)$ in general number of spatial dimensions, obeying the
quasiperiodicity conditions
\begin{equation}
\psi (t,x^{1},\ldots ,x^{p},\ldots ,x^{l}+L_{l},\ldots ,x^{D})=e^{i\alpha
_{l}}\psi (t,x^{1},\ldots ,x^{p},\ldots ,x^{l},\ldots ,x^{D}),  \label{Qperf}
\end{equation}%
with constant phases $\alpha _{l}$, are presented in \cite%
{Bell10,Bell13dS,Bell17AdS} for LM, LdS and LAdS geometries, respectively.
The formulas from these references for the fermionic current density along
the $r$th compact dimension are presented in the combined form
\begin{equation}
\left\langle j_{\mathrm{(p)}}^{r}\right\rangle _{\mathrm{(f)}}=-\frac{NeL_{%
\mathrm{(p)}r}}{\left( 2\pi \right) ^{\frac{D+1}{2}}a^{D+1}}\sum_{\mathbf{n}%
_{q}}n_{r}\sin \left( \mathbf{n}_{q}\cdot \mathbf{\tilde{\alpha}}_{q}\right)
F_{D}^{\mathrm{(f)}}(am,g(\mathbf{L}_{\mathrm{(p)}q}/a,\mathbf{n}_{q})),
\label{jrcf}
\end{equation}%
with the same notations as in (\ref{jrc}). Here, $N=2^{\left[ \frac{D+1}{2}%
\right] }$ ($[x]$ stands for the integer part of $x$) is the number of
spinor components for the Dirac field realizing the irreducible
representation of the Clifford algebra. The functions $F_{D}^{\mathrm{(f)}%
}(am,x)$ in (\ref{jrcf}) are defined by
\begin{eqnarray}
F_{D}^{\mathrm{(f)}}(am,x) &=&F_{D}(am,x),\mathrm{\;for\;LM},  \notag \\
F_{D}^{\mathrm{(f)}}(am,x) &=&\frac{1}{2}{\mathrm{Re}}\left[ p_{i\alpha m}^{-%
\frac{D+1}{2}}\left( x^{2}/2-1\right) \right] ,\mathrm{\;for\;LdS},  \notag
\\
F_{D}^{\mathrm{(f)}}(am,x) &=&\frac{1}{2}\left[ q_{am}^{\frac{D+1}{2}}\left(
x^{2}/2+1\right) +q_{am-1}^{\frac{D+1}{2}}\left( x^{2}/2+1\right) \right] ,%
\mathrm{\;for\;LAdS}.  \label{FD2f}
\end{eqnarray}%
The replacement $2\pi \tilde{\alpha}_{l}\rightarrow -\tilde{\alpha}_{l}$ in
the expression for LM bulk, compared to the one given in \cite{Bell10}, is
related to different notations of the constants in the quasiperiodicity
conditions (see also the comment in \cite{Bell17AdS}). The applications of (%
\ref{jrcf}), with $D=2$, in cylindrical nanotubes, described in terms of the
effective Dirac theory, have been discussed in \cite{Bell10,Bell17AdS}.

As it is seen from (\ref{FD2f}), assuming the same masses and phases in the
periodicity conditions for scalar and Dirac fields, the relation $\langle
j^{l}\rangle _{\mathrm{(f)LM}}=-(N/2)\langle j^{l}\rangle _{\mathrm{LM}}$ is
obtained for the corresponding current densities in LM\ bulk. In particular,
in supersymmetric models with the same number of scalar and spinor degrees
of freedom the total vacuum current vanishes. That is not the case for LdS
and LAdS geometries. In even number of spatial dimensions $D$, the Clifford
algebra has two inequivalent representations with two different sets of the
Dirac matrices. As it has been discussed in \cite{Bell17AdS}, the vacuum
current densities coincide for the fields realizing those representations if
the corresponding masses and the periodicity conditions are the same. More
details of the properties for fermionic currents in LM, LdS and LAdS
geometries with toroidal compact dimensions will be reviewed elsewhere.

\section{Conclusion}

\label{sec:Conc}

In the present paper we have discussed the features of the vacuum currents
in field-theoretical models formulated in background of spacetimes with
compact dimensions. Three cases of background geometries are considered: LM,
LdS and LAdS. In the decompactification limit they correspond to maximally
symmetric solutions of the Einstein field equations in $(D+1)$-dimensional
spacetime with zero, positive and negative cosmological constants,
respectively. The toroidal compactification of a part of spatial dimensions
does not change the local geometrical characteristics and the high symmetry
allows to find closed analytic expressions for the vacuum currents along
compact dimensions. For an external gauge field we have taken the simplest
configuration with a constant gauge field. Though the corresponding magnetic
field is zero, because of the nontrivial topology respective vector
potential gives rise to Aharonov-Bohm like effect on the vacuum
characteristics. By a gauge transformation the gauge field potential is
reinterpreted in terms of the phases in the periodicity conditions on the
field operator along compact dimensions. The quasiperiodicity conditions
with nontrivial phases break the reflection symmetry along the respective
direction and, as a consequence, the contributions of the left and right
moving modes of the vacuum fluctuations of the quantum field do not
compensate each other. As a result a net current appears that is the analog
of the persistent currents in mesoscopic metallic rings.

The combined expression for the current density along the $r$th compact
dimension, valid for all three background geometries, is given by the
formula (\ref{jrc}). The information on specific geometry is encoded in the
function $F_{D}(am,x)$ defined by (\ref{FD}). The component of the current
density $\left\langle j_{\mathrm{(p)}}^{r}\right\rangle $ is an odd periodic
function of the phase $\tilde{\alpha}_{r}$ with the period $2\pi $ and an
even periodic function of the remaining phases $\tilde{\alpha}_{l}$, $l\neq
r $, with the same period. This periodicity is also interpreted as
periodicity in terms of the magnetic flux enclosed by compact dimensions. In
this interpretation the period is equal to the flux quantum. For curved
backgrounds the current density depends on the lengths of the compact
dimensions and on the coordinates (temporal $\tau $ and spatial $z$
coordinates for LdS and LAdS, respectively) in the form of the proper
lengths $L_{\mathrm{(p)}l}$. This feature is a consequence of the maximal
symmetry of dS and AdS spacetimes. For a conformally coupled massless scalar
field the current densities in LM and LdS spacetimes are connected by the
standard relation (\ref{jrdSm0}). For LAdS geometry one has a conformal
relation with the current density in LM spacetime (given by (\ref{jrMm0b}))
with an additions planar boundary perpendicular to one of the uncompact
dimensions. The boundary in LM spacetime with Dirichlet boundary condition
on the scalar field operator is the conformal image of the AdS boundary.

For LdS and LAdS bulks and for small values of the length of compact
dimension the mode sum of the component of the current density along that
dimension is dominated by the contribution of the vacuum fluctuations with
wavelengths smaller than the curvature radius. The influence of the
gravitational field on those modes is weak and the leading term in the
respective expansion, given by (\ref{jrpsm}), coincides with that for LM\
bulk with the length of the compact dimension replaced by the proper lengths
for LdS and LAdS geometries. The leading term presents the current density
for a massless field in $(D+1)$-dimensional LM\ spacetime with a single
compact dimension of the length $L_{\mathrm{(p)}r}$. For small values of the
length of the $l$th compact dimension, the behavior of the current density
along the $r$th dimension, $r\neq l$, essentially differs for zero and
nonzero values of the phase $|\tilde{\alpha}_{l}|\leq 1/2$. In the first
case, $\tilde{\alpha}_{l}=0$, the dominant contribution to the current
density $\left\langle j_{\mathrm{(p)}}^{r}\right\rangle $ comes from the
zero mode $n_{l}=0$ and the leading term in the expansion of the product $L_{%
\mathrm{(p)}l}\left\langle j_{\mathrm{(p)}}^{r}\right\rangle $ coincides
with the corresponding current density in $D$-dimensional spacetime with
spatial coordinates $(x^{1},\ldots ,x^{l-1},x^{l+1},\ldots ,x^{D})$. For $%
\tilde{\alpha}_{l}\neq 0$ the zero mode with respect to the $l$th dimension
is absent and the component $\left\langle j_{\mathrm{(p)}}^{r}\right\rangle $
decays like $\exp (-|\tilde{\alpha}_{l}|L_{r}/L_{l})$.

The effect of the spacetime curvature on the current density is essential
for lengths of compact dimensions of the order or larger compared with the
curvature radius. For large values of the length of the $r$th compact
dimension the asymptotic of the current $\left\langle j_{\mathrm{(p)}%
}^{r}\right\rangle $ is completely different for the cases $\omega _{0r}=0$
and $\omega _{0r}\neq 0$ with $\omega _{0r}=\sqrt{k_{\mathbf{n}%
_{q-1}}^{(0)2}+m^{2}}$ and $k_{\mathbf{n}_{q-1}}^{(0)2}$ defined by (\ref%
{k0r}). For $\omega _{0r}=0$, corresponding to a massless field with zero
phases $\tilde{\alpha}_{l}$, $l\neq r$, the leading term for the LM\ bulk is
given by (\ref{jrMlLm0}). Multiplied by $V_{q}/L_{r}$, that expression gives
the current density for a massless field in $(p+2)$-dimensional LM spacetime
with a single compact dimension $x^{r}$. For $\omega _{0r}\neq 0$ the large $%
L_{r}$ asymptotic is described by (\ref{jrT0large}) and the current density
in LM bulk is exponentially suppressed. For LdS and LAdS background
geometries and for $\tilde{\alpha}_{l}=0$, $l\neq r$, the leading term in
the large $L_{\mathrm{(p)}r}$ asymptotic is given by the right-hand side of (%
\ref{jrLargeLp}) with $\nu >0$ for LdS bulk. This shows that the
gravitational field essentially changes the behavior of the current density
for large lengths of compact dimensions: instead of the exponential
suppression in LM\ bulk for a massive field, for LdS and LAdS geometries the
fall-off of the current density follows a power law. For LdS background and
for imaginary values of the parameter $\nu $ the behavior of the current
density is described by (\ref{jrLargLpim2}). In this case the decay with
respect to $L_{r}$ is oscillatory with the amplitude decreasing as $%
1/L_{r}^{p+1}$. In the case when at least one of the phases $\tilde{\alpha}%
_{l}=0$, $l\neq r$, differs from zero, the asymptotic behavior of the
current density $\left\langle j_{\mathrm{(p)}}^{r}\right\rangle $ is given
by (\ref{jrpLargeLp2}) for LAdS and for LdS in the range $\nu >0$ with an
exponential decay. For LdS bulk and imaginary $\nu $ the decay is
oscillatory (see (\ref{jrLargLpim2})).

The current density along compact dimensions is a source of magnetic fields
having components in the noncompact subspace. In spatial dimensions $D>3$
the magnetic field is a spatial tensor of rank $D-2$ which can be found by
solving Maxwell's $(D+1)$-dimensional semiclassical equations with the VEV
of the current density as a source. That would be an interesting application
of the results described in the present paper. Note that several mechanisms
for the generation of the seeds for cosmological magnetic fields in
higher-dimensional models have been discussed in the literature (see, for
example, \cite{Kand11,Camp13}).

\section*{Acknowledgments}

I am grateful to Stefano Bellucci, Eugenio Ramos Bezerra de Mello and Valeri
Vardanyan for fruitful collaboration. The work was supported by the grant
No. 21AG-1C047 of the Higher Education and Science Committee of the Ministry
of Education, Science, Culture and Sport RA and by the ANSEF grant
23AN:PS-hepth-2889.

\appendix

\section{Properties of the Functions in the Expressions for the Currents}

\label{sec:App1}

We have seen that the vacuum currents along compact dimensions in LdS and
LAdS geometries are expressed in terms of the functions (\ref{pmu}) and (\ref%
{qmu}). Here the properties of those functions are considered. First of all
we note that they are expressed in terms of the hypergeometric function as%
\begin{eqnarray}
p_{\nu -\frac{1}{2}}^{-\mu }(u) &=&\frac{\Gamma (\mu +\frac{1}{2}-\nu
)\Gamma (\mu +\frac{1}{2}+\nu )}{2^{\mu }\Gamma \left( \mu +1\right) }%
F\left( \mu +\frac{1}{2}-\nu ,\mu +\frac{1}{2}+\nu ;\mu +1;\frac{1-u}{2}%
\right) ,  \notag \\
q_{\nu -\frac{1}{2}}^{\mu }(u) &=&\frac{\sqrt{\pi }\Gamma (\mu +\nu +\frac{1%
}{2})}{2^{\nu +\frac{1}{2}}\Gamma (\nu +1)u^{\mu +\nu +\frac{1}{2}}}F\left(
\frac{\mu +\nu +\frac{3}{2}}{2},\frac{\mu +\nu +\frac{1}{2}}{2};\nu +1;\frac{%
1}{u^{2}}\right) .  \label{qu1}
\end{eqnarray}%
By using the recurrence relations for the associated Legendre functions one
can see that $\partial _{u}p_{\alpha }^{-\mu }(u)=-p_{\alpha }^{-\mu -1}(u)$
and $\partial _{u}q_{\alpha }^{\mu }(u)=-q_{\alpha }^{\mu +1}(u)$. From here
we get the relations%
\begin{equation}
p_{\alpha }^{-\mu -n}(u)=(-1)^{n}\partial _{u}^{n}p_{\alpha }^{-\mu
}(u),\;q_{\alpha }^{\mu +n}(u)=(-1)^{n}\partial _{u}^{n}q_{\alpha }^{\mu
}(u),  \label{pqrel}
\end{equation}%
with $n=1,2,\ldots $.

In the physical problems under consideration, depending on the spatial
dimension, the order $\mu $ for the functions $p_{\alpha }^{-\mu }(u)$ and $%
q_{\alpha }^{\mu }(u)$ is a positive integer or half-integer. By making use
of (\ref{pqrel}) these functions are expressed in terms of $p_{\alpha
}^{0}(u)$ and $q_{\alpha }^{0}(u)$ or $p_{\alpha }^{-1/2}(u)$ and $q_{\alpha
}^{1/2}(u)$. Employing the corresponding expressions for the functions $%
P_{\alpha }^{-1/2}(u)$ and $Q_{\alpha }^{1/2}(u)$ from \cite{Abra} one
obtains%
\begin{eqnarray}
p_{\nu -1/2}^{-1/2}(\cosh \zeta ) &=&\frac{\sqrt{2\pi }\sinh \left( \nu
\zeta \right) }{\sin \left( \pi \nu \right) \sinh \zeta },  \notag \\
p_{\nu -1/2}^{-1/2}(\cos \theta ) &=&\frac{\sqrt{2\pi }\sin \left( \nu
\theta \right) }{\sin \left( \pi \nu \right) \sin \theta },  \notag \\
q_{\nu -1/2}^{1/2}(\cosh \zeta ) &=&\sqrt{\frac{\pi }{2}}\frac{e^{-\nu \zeta
}}{\sinh \zeta }.  \label{pq05}
\end{eqnarray}%
Combining these expressions with the relations (\ref{pqrel}) for even values
of the spatial dimension $D$ we get
\begin{equation}
F_{D}(am,x)=\sqrt{\frac{\pi }{2}}(-1)^{\frac{D}{2}}\left( \frac{\partial
_{\xi }}{\sinh \xi }\right) ^{\frac{D}{2}}\frac{e^{-\nu _{+}\xi }}{\sinh \xi
},\;x=2\sinh \left( \xi /2\right) ,  \label{FDadsEv}
\end{equation}%
in LAdS geometry and%
\begin{eqnarray}
F_{D}(am,x) &=&\frac{(-1)^{\frac{D}{2}}\sqrt{\pi /2}}{\sin \left( \pi \nu
\right) }\left( \frac{\partial _{\xi }}{\sinh \xi }\right) ^{\frac{D}{2}}%
\frac{\sinh \left( \nu \xi \right) }{\sinh \xi },\;x=2\cosh \left( \xi
/2\right) ,  \notag \\
F_{D}(am,x) &=&\frac{\sqrt{\pi /2}}{\sin \left( \pi \nu \right) }\left(
\frac{\partial _{\theta }}{\sin \theta }\right) ^{\frac{D}{2}}\frac{\sin
\left( \nu \theta \right) }{\sin \theta },\;x=2\cos \left( \theta /2\right) ,
\label{FDdSEv}
\end{eqnarray}%
for LdS bulk. For odd values of $D$ one has
\begin{eqnarray}
F_{D}(am,x) &=&\frac{1}{2}(-1)^{\frac{D+1}{2}}\Gamma \left( \frac{1}{2}-\nu
\right) \Gamma \left( \nu +\frac{1}{2}\right) \partial _{u}^{\frac{D+1}{2}%
}P_{\nu -\frac{1}{2}}(u),\mathrm{\;for\;LdS},  \notag \\
F_{D}(am,x) &=&(-1)^{\frac{D+1}{2}}\partial _{u}^{\frac{D+1}{2}}Q_{\nu _{+}-%
\frac{1}{2}}(u),\mathrm{\;for\;LAdS},  \label{FDod}
\end{eqnarray}%
where $u=x^{2}/2-1$ for LdS and $u=x^{2}/2+1$ for LAdS.

In order to find the behavior of the functions $p_{\nu -1/2}^{-\mu }(u)$ and
$q_{\nu -1/2}^{\mu }(u)$ for large values of $u$ we use the corresponding
asymptotics for the associated Legendre functions (see, for example, \cite%
{Olve10}). The leading order terms read
\begin{eqnarray}
p_{\nu -\frac{1}{2}}^{-\mu }(u) &\sim &\frac{2^{\nu -\frac{1}{2}}}{\sqrt{\pi
}}\frac{\Gamma \left( \nu \right) \Gamma \left( \mu +\frac{1}{2}-\nu \right)
}{u^{\mu -\nu +\frac{1}{2}}},\;\mathrm{Re}\,\nu >0,\;\mu +\nu \neq
-1,-2,\ldots ,  \notag \\
p_{-\frac{1}{2}}^{-\mu }(u) &\sim &\sqrt{\frac{2}{\pi }}\Gamma \left( \mu +%
\frac{1}{2}\right) \frac{\ln u}{u^{\mu +\frac{1}{2}}},\;\mu \neq -\frac{1}{2}%
,-\frac{3}{2},\ldots ,  \notag \\
q_{\nu -\frac{1}{2}}^{\mu }(u) &\sim &\frac{\sqrt{\pi }\Gamma \left( \mu
+\nu +\frac{1}{2}\right) }{2^{\nu +\frac{1}{2}}\Gamma \left( \nu +1\right)
u^{\mu +\nu +\frac{1}{2}}},\;\nu \neq -\frac{3}{2},-\frac{5}{2},\ldots
\label{pqas}
\end{eqnarray}%
The asymptotic for $p_{\nu -\frac{1}{2}}^{-\mu }(u)$ in the case of purely
imaginary values of $\nu $, $\nu =i|\nu |$, is obtained by using the
relation \cite{Olve10}
\begin{equation}
P_{i|\nu |-\frac{1}{2}}^{-\mu }(u)=ie^{-i\mu \pi i}\frac{Q_{i|\nu |-\frac{1}{%
2}}^{\mu }(u)-Q_{-i|\nu |-\frac{1}{2}}^{\mu }(u)}{\sinh \left( |\nu |\pi
\right) |\Gamma \left( \mu +\frac{1}{2}+i|\nu |\right) |^{2}},  \label{PQrel}
\end{equation}%
and the corresponding asymptotic for $Q_{i|\nu |-\frac{1}{2}}^{\mu }(u)$.
This gives%
\begin{equation}
p_{i|\nu |-\frac{1}{2}}^{-\mu }(u)\sim \frac{\sqrt{2/\pi }}{u^{\mu +\frac{1}{%
2}}}\,\mathrm{Re}\left[ \Gamma \left( \mu +\frac{1}{2}-i|\nu |\right) \frac{%
\Gamma \left( i|\nu |\right) }{\left( 2u\right) ^{i|\nu |}}\right] .
\label{pas}
\end{equation}%
These asymptotic formulas have been used in the main text to study the
behavior of the current density in asymptotic regions of the parameters.

\end{document}